\documentclass[sn-mathphys-num]{sn-jnl}

\usepackage{graphicx}%
\usepackage{multirow}%
\usepackage{amsmath,amssymb,amsfonts}%
\usepackage{amsthm}%
\usepackage{mathrsfs}%
\usepackage[title]{appendix}%
\usepackage{xcolor}%
\usepackage{textcomp}%
\usepackage{manyfoot}%
\usepackage{booktabs}%
\usepackage{algorithm}%
\usepackage{algorithmicx}%
\usepackage{algpseudocode}%
\usepackage{listings}%
\usepackage{todonotes}%

\theoremstyle{thmstyleone}%
\theoremstyle{thmstyletwo}%

\theoremstyle{thmstylethree}%

\usepackage{color,soul}
\usepackage[hang,bottom]{footmisc}
\usepackage{amsmath} 
\usepackage{framed} 
\usepackage{multicol} 
\usepackage{nomencl} 
\makenomenclature
\renewcommand*\nompreamble{\begin{multicols}{2}}
\renewcommand*\nompostamble{\end{multicols}}
\usepackage{siunitx} 
\usepackage{color}
\usepackage{tabularray}
\usepackage{array}
\usepackage{booktabs}
\usepackage{longtable}
\usepackage{subcaption}
\usepackage{array}
\usepackage{pdflscape}
\usepackage{longtable}
\usepackage{booktabs}
\usepackage{hyperref}
\usepackage{bbding}
\usepackage{pifont}

\usepackage[utf8]{inputenc}
\usepackage{pdflscape}

\usepackage{xurl}

\begin{document}

\title[Article Title]{Gaussian Models to Non-Gaussian Realms of Quantum Photonic Simulators}


\author*[1]{\fnm{Dennis Delali Kwesi} \sur{Wayo}}\email{dennis.wayo@nu.edu.kz}

\author[2]{\fnm{Rodrigo Alves} \sur{Dias}}\email{diasrodri@gmail.com}

\author*[3]{\fnm{Masoud Darvish} \sur{Ganji}}\email{ganji\_md@yahoo.com}

\author[4]{\fnm{Camila Martins} \sur{Saporetti}}\email{camila.saporetti@iprj.uerj.br}

\author*[5]{\fnm{Leonardo} \sur{Goliatt}}\email{leonardo.goliatt@ufjf.br}

\affil*[1]{\orgdiv{Faculty of Chemical and Process Engineering Technology}, \orgname{Universiti Malaysia Pahang Al-Sultan Abdullah}, \orgaddress{\city{Kuantan}, \postcode{26300}, \country{Malaysia}}}

\affil[2]{\orgdiv{Department of Physics}, \orgname{Federal University of Juiz de Fora}, \orgaddress{\city{Juiz de Fora}, \postcode{36036-900},\country{Brazil}}}

\affil*[3]{\orgdiv{Division of Carbon Neutrality and Digitalization}, \orgname{Korea Institute of Ceramic Engineering and Technology (KICET)}, \orgaddress{\city{Jinju}, \postcode{52851}, \country{Republic of Korea}}}

\affil[4]{\orgdiv{Department of Computational Modeling}, \orgname{, Polytechnic Institute, Rio de Janeiro State University}, \orgaddress{\city{Nova Friburgo}, \postcode{28625-570},\country{Brazil}}}

\affil*[5]{\orgdiv{Department of Computational and Applied Mechanics}, \orgname{Federal University of Juiz de Fora}, \orgaddress{\city{Juiz de Fora}, \postcode{36036-900},\country{Brazil}}}

\abstract{Quantum photonic simulators have emerged as indispensable tools for modeling and optimizing quantum photonic circuits, bridging the gap between theoretical models and experimental implementations. This review explores the landscape of photonic quantum simulation, focusing on the transition from Gaussian to non-Gaussian models and the computational challenges associated with simulating large-scale photonic systems. Gaussian states and operations, which enable efficient simulations through covariance matrices and phase-space representations, serve as the foundation for photonic quantum computing. However, non-Gaussian states crucial for universal quantum computation introduce significant computational complexity, requiring advanced numerical techniques such as tensor networks and high-performance GPU acceleration. We evaluate the leading photonic quantum simulators, including Strawberry Fields, Piquasso, QuTiP SimulaQron, Perceval, and QuantumOPtics.jl analyzing their capabilities in handling continuous-variable (CV) and discrete-variable (DV) quantum systems. Special attention is given to hardware-accelerated methods, including GPU-based tensor network approaches, machine learning integration, and hybrid quantum-classical workflows. Furthermore, we investigate noise modeling techniques, such as photon loss and dark counts, and their impact on simulation accuracy. As photonic quantum computing moves toward practical implementations, advancements in high-performance computing (HPC) architectures, such as tensor processing units (TPUs) and system-on-a-chip (SoC) solutions, are accelerating the field. This review highlights emerging trends, challenges, and future directions for developing scalable and efficient photonic quantum simulators.}



\maketitle
\section{Introduction}

\begin{figure}[H]
\centering
\includegraphics[width=0.5\textwidth]{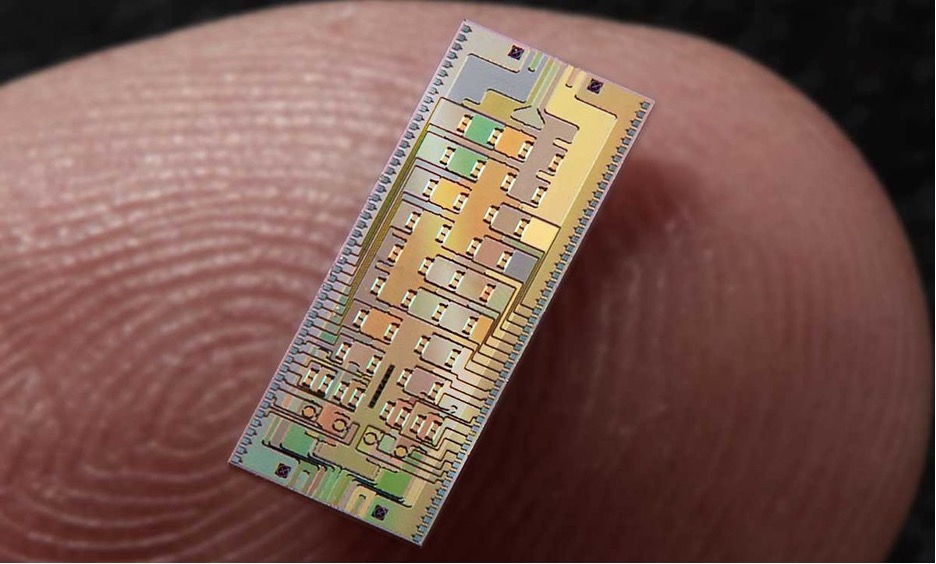}
\caption{This image showcases Xanadu's X8 photonic quantum computing chip, a programmable, scalable quantum processor leveraging continuous-variable (CV) quantum photonics. Unlike superconducting or trapped-ion quantum computers, which require extreme cryogenic cooling, photonic quantum computers operate at room temperature, making them easier to integrate into existing fiber-optic infrastructure. The X8 chip, measuring 4 mm $\times$ 10 mm, acts as an 8-qubit processor, using infrared laser pulses, squeezed states, beam splitters, and phase shifters to perform computations. Xanadu provides cloud access to the chip via Strawberry Fields and PennyLane, enabling remote quantum photonic computing, adapted from IEEE Spectrum: Race to Hundreds of Photonic Qubits \citet{choi2021race}}  \label{fig1}
\end{figure}

\begin{figure}[H]
\centering
\includegraphics[width=0.7\textwidth]{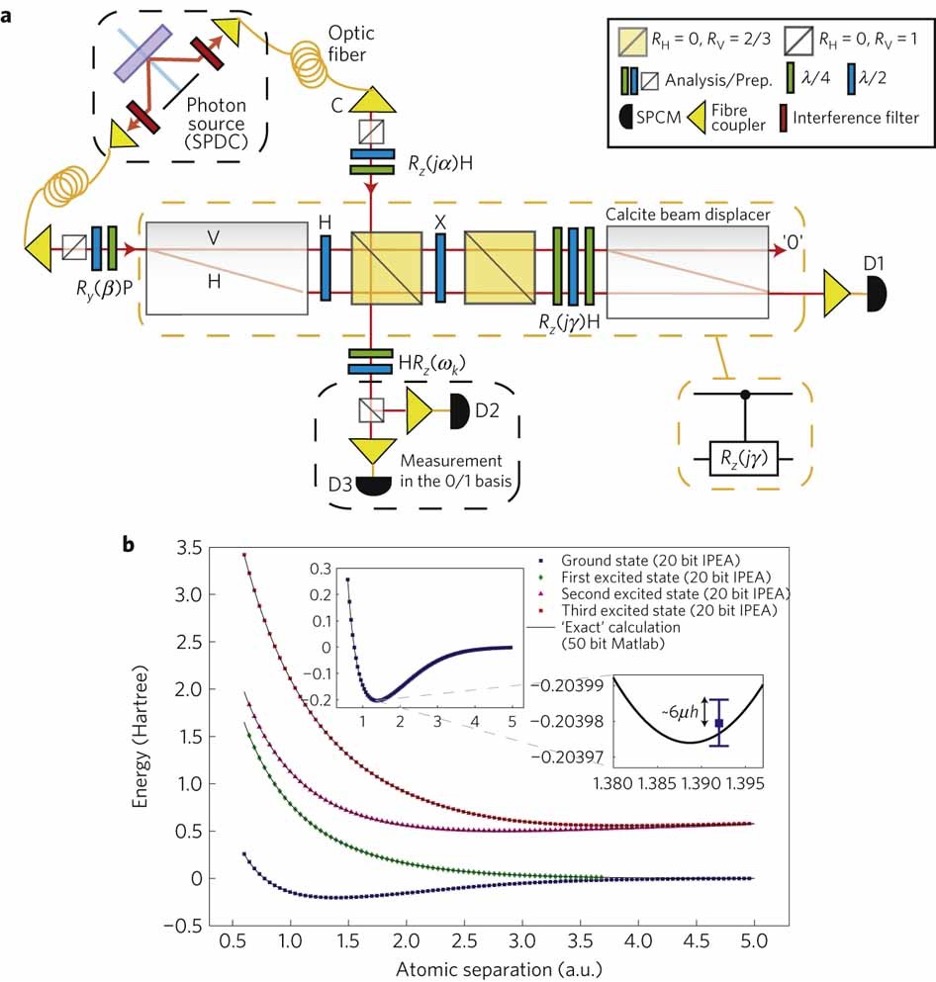}
\caption{Aspuru-Guzik-Walther demonstrated in the above figure a photonic quantum simulation used to calculate molecular energy levels. (a) depicts an optical quantum circuit, where single photons generated via spontaneous parametric down-conversion (SPDC) are manipulated using beam splitters, phase shifters, and wave plates to implement quantum gates. The circuit encodes quantum computations, and measurements are taken in the 0/1 basis using detectors. (b) presents the simulated energy levels (Hartree) as a function of atomic separation, comparing quantum results (20-bit IPEA) with classical calculations (50-bit Matlab). This demonstrates the potential of photonic quantum computing in quantum chemistry and molecular modeling, adapted with permission from \citet{aspuru2012photonic}}
    \label{fig2}
\end{figure}

Photonic quantum computing (PQC) is rapidly evolving as a transformative technology with the potential to solve problems that are intractable for classical systems \cite{feynman1982simulating}. Among the various quantum computing platforms, as demonstrated by Xanadu's X8 chip in  Figure \ref{fig1} photonic quantum systems \cite{choi2021race} stand out due to their inherent scalability, high coherence times, and room-temperature operation. By encoding information in quantum states of light as presented in Figure \ref{fig2}, photons’ polarization, phase, or spatial modes, photonic systems offer a promising pathway toward practical and fault-tolerant quantum computation. This unique platform, however, presents its own set of challenges, especially in the simulation and design of photonic quantum circuits.

The simulation of quantum photonic systems is essential for researchers and engineers aiming to design, validate, and optimize photonic circuits before implementing them in hardware \cite{aspuru2012photonic}. Simulation tools serve as a bridge between theoretical proposals and experimental realizations, allowing researchers to model complex interactions, predict outcomes, and explore novel quantum algorithms \cite{killoran2019strawberry, QuTiP2015, georgescu2014quantum, kolarovszki2024piquasso, buluta2009quantum, fauseweh2024quantum}. These tools are particularly critical in the photonic domain, where the interplay of Gaussian and non-Gaussian states determines the computational power of a given system \cite{su2019conversion}. Gaussian models \cite{wang2007quantum}, characterized by operations like squeezing and displacement, are computationally efficient but limited in their ability to achieve universal quantum computation. Non-Gaussian states \cite{walschaers2021non, filip2013gaussian}, introduced via photon-number-resolving detectors or Kerr interactions, are necessary for universality but significantly increase the complexity of simulations.

The shift from Gaussian to non-Gaussian realms in quantum photonic simulators represents a critical challenge in the field. Gaussian states, owing to their mathematical simplicity, can be efficiently simulated using tools like covariance matrices and phase-space representations. These methods exploit the properties of Gaussian distributions \cite{blinnikov1998expansions, carreira2000mode, chhikara1988inverse}, ensuring scalability even for large systems. Non-Gaussian states, on the other hand, disrupt this simplicity, requiring full quantum state representations, such as Fock states or density matrices \cite{lachman2022quantum}. This transition introduces exponential growth in computational requirements, making it imperative to develop new techniques and tools that balance accuracy and efficiency.

Several quantum photonic simulators have been developed to address these needs, each with unique strengths and limitations. Xanadu’s Strawberry Fields, for instance, has become a cornerstone in the field, offering comprehensive support for continuous-variable (CV) quantum computing \cite{jha2024continuous,matsuura2024continuous,haque2024continuous} and a Python-based API that integrates seamlessly with machine learning frameworks like PennyLane \cite{bergholm2018pennylane}. Similarly, Ansys lumerical \cite{puspaduhita2023optical} provides robust tools for modeling the physical properties of photonic components, enabling circuit-level design and optimization. Despite these advancements, significant gaps remain. Many simulators struggle \cite{zohar2022quantum,bauer2023quantum} with scalability when modeling large circuits with non-Gaussian elements or fail to integrate effectively with hybrid quantum-classical workflows. Moreover, the steep learning curve of some tools can deter researchers and educators who lack extensive computational backgrounds.

Beyond their role in research, photonic quantum simulators also serve as educational tools. As quantum technologies become more prevalent, there is a growing need to train a new generation of scientists and engineers in the principles of quantum photonics. Simulators with intuitive interfaces and comprehensive visualization capabilities can demystify complex quantum phenomena, making them accessible to a broader audience. Such tools are particularly valuable in academic settings, where they can complement theoretical coursework with hands-on experimentation in a virtual environment.

The motivation for this review lies in the need to systematically examine the current landscape of quantum photonic simulators and identify the pathways for future development. This work surveys the foundational principles of photonic quantum computing, evaluates the capabilities of existing simulation tools, and highlights the computational techniques that enable scalable and accurate simulations. A comparative analysis of tools, such as Strawberry Fields, Ansys lumerical, and emerging platforms, reveals the progress made and the challenges that persist. The review also explores advancements in hardware acceleration, tensor networks, and hybrid quantum-classical workflows, all of which are instrumental in addressing the limitations of current simulators.

Central to this review is the idea of democratizing access to quantum photonic simulation tools. Open-source initiatives have already transformed other domains of quantum computing by fostering collaboration and innovation. Extending this philosophy to photonic quantum simulators could accelerate the adoption of photonic quantum technologies, enabling researchers, educators, and developers from diverse backgrounds to contribute to the field. By making simulation tools more accessible and user-friendly, the community can unlock new opportunities for innovation and discovery.

The structure of this review is designed to provide a comprehensive understanding of the field. Following this introduction, the foundational principles of quantum photonic simulation are discussed, including Gaussian and non-Gaussian models, continuous-variable frameworks, and noise modeling. The subsequent section surveys the major quantum photonic simulators, analyzing their features, limitations, and applications. A dedicated section explores the computational techniques that drive advancements in scalability and performance, such as tensor networks, GPU acceleration, and machine learning. The review concludes with a discussion of challenges and open problems, as well as recommendations for future directions.

Quantum photonic simulators are indispensable tools in the quest for scalable quantum computing. They enable researchers to design and optimize circuits, explore quantum algorithms, and bridge the gap between theory and experiment. However, as the field transitions from Gaussian to non-Gaussian realms, the limitations of existing tools become increasingly apparent. This review aims to illuminate the current state of quantum photonic simulators, identify the gaps that need to be addressed, and chart a path forward for the development of next-generation tools that are versatile, scalable, and accessible to all.

\section{Foundations of Quantum Photonic Simulation}

\subsection{Photonic Quantum Computing Basics}

Quantum photonic computing \cite{wayo2025linear} relies on the quantum states of light to encode, manipulate, and process information. Photons, as the carriers of information, possess unique properties such as coherence, high-speed transmission, and resistance to thermal noise, making them ideal candidates for quantum systems. Photonic quantum computing operates on two primary paradigms: continuous-variable (CV) \cite{anai2024continuous, bangar2024continuous, iosue2024continuous, zhang2024continuous,shaikh2024focked} and discrete-variable (DV) systems \cite{kish2024comparison}. While CV systems encode quantum information in continuous properties such as amplitude and phase, DV systems use discrete states such as the presence or absence of a photon in a given mode. These paradigms form the foundation for simulating photonic quantum systems, with each presenting distinct computational and experimental challenges.

In CV systems, operations such as squeezing, displacement, and rotation can be represented using Gaussian models, which allow for efficient simulations using covariance matrices and phase-space methods. DV systems, on the other hand, often require non-Gaussian states, such as single-photon sources or photon-number-resolving detectors, to achieve computational universality. The interplay between these paradigms necessitates simulation tools to model hybrid systems, combining the strengths of CV and DV approaches to explore new computational possibilities.

\subsection{Gaussian Models}

\begin{figure}[H]
\centering
\includegraphics[width=0.8\textwidth]{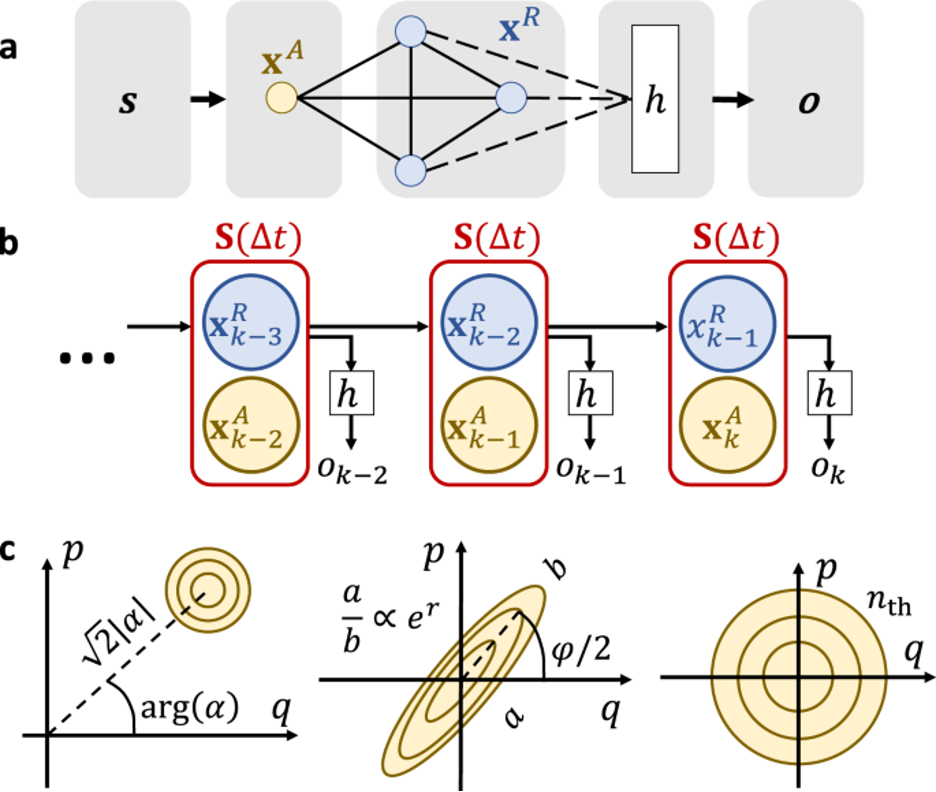}
\caption{A Gaussian model framework used in quantum photonic systems. (a) depicts a neural network-like structure, where an input state \(s\) is processed through layers of variables x\(^A\) and x\(^R \), with a transformation function \(h\) leading to an output \(o\). (b) represents a sequential evolution process, where Gaussian quantum states evolve over discrete time steps \(S(\Delta t)\). (c) visualizes phase-space representations of Gaussian states, showing coherent state displacement, squeezed states, and thermal states in \( p-q \) phase space. This model is essential for quantum optics and machine learning applications, adapted with permission from \citet{nokkala2021gaussian}}
    \label{fig3}
\end{figure}

Gaussian models as presented by Nokkala in Figure \ref{fig3} forms a keystone in quantum photonic simulations due to their mathematical simplicity, scalability, and foundational role in continuous-variable quantum computing \cite{menicucci2006universal}. These models describe quantum states and operations whose representations in phase space are fully characterized by Gaussian distributions.

Gaussian states, such as coherent states, squeezed states, and thermal states, are defined by their first and second moments \cite{feng2024evsplitting}:
    \begin{enumerate}
        \item [1.] First Moments ($\mathbf{d}$): The mean values of the quadrature operators, represented as a vector \cite{souza2024classes, brandao2022qugit}:
    \end{enumerate}

\begin{equation}
    \mathbf{d} = \langle \mathbf{R} \rangle, \quad \mathbf{R} = \begin{pmatrix} \hat{x}_1, \hat{p}_1, \dots, \hat{x}_n, \hat{p}_n \end{pmatrix}^T,
\end{equation}

where $\hat{x}_j = \frac{\hat{a}_j + \hat{a}_j^\dagger}{\sqrt{2}}$ and $\hat{p}_j = i\frac{\hat{a}_j^\dagger - \hat{a}_j}{\sqrt{2}}$ are the quadrature operators for the j-th mode.

    \begin{enumerate}
        \item [2.] Second Moments ($\sigma$): The covariance matrix \cite{cariolaro2022gaussian,garcia2020covariance,werner2001bound}, defined as:
    \end{enumerate}

\begin{equation}
    \sigma_{ij} = \frac{1}{2} \langle \hat{R}_i \hat{R}_j + \hat{R}_j \hat{R}_i \rangle - \langle \hat{R}_i \rangle \langle \hat{R}_j \rangle.
\end{equation}

Gaussian operations \cite{strawberryfields_states}, such as displacement, squeezing, and beam splitting, preserve the Gaussian nature of states. These operations \cite{miller2025curvature} are represented as symplectic transformations $\mathbf{S}$ acting on the covariance matrix $\sigma$ and mean vector $\mathbf{d}$:

\begin{equation}
    \sigma \to \mathbf{S} \sigma \mathbf{S}^T, \quad \mathbf{d} \to \mathbf{S} \mathbf{d}.
\end{equation}

\subsubsection{Wigner Function Representation}

The Wigner function \cite{wootters1987wigner, veitch2013efficient, weinbub2018recent, feng2024connecting} $W(\mathbf{r})$ for a Gaussian state is expressed as:

\begin{equation}
    W(\mathbf{r}) = \frac{\exp\left(-\frac{1}{2} (\mathbf{r} - \mathbf{d})^T \sigma^{-1} (\mathbf{r} - \mathbf{d}) \right)}{(2\pi)^n \sqrt{\det \sigma}},
\end{equation}

where $\mathbf{r}$ represents the phase-space variables.

\subsubsection{Symplectic Framework for Gaussian Systems}

The symplectic structure of phase space facilitates efficient simulations of Gaussian systems. The evolution of a Gaussian state \cite{linowski2022dissipative} is governed by symplectic matrices $\mathbf{S}$ satisfying:

\begin{equation}
    \mathbf{S} \Omega \mathbf{S}^T = \Omega, \quad \Omega = \bigoplus_{j=1}^n \begin{pmatrix} 0 & 1 \\ -1 & 0 \end{pmatrix}.
\end{equation}

For an n-mode system, the Hamiltonian governing Gaussian dynamics is at most quadratic in the quadrature operators:
\begin{equation}
    \hat{H} = \frac{1}{2} \mathbf{R}^T \mathbf{H} \mathbf{R} + \mathbf{R}^T \mathbf{h} + h_0,
\end{equation}
where $\mathbf{H}$ is a symmetric 2n $\times$ 2n matrix, $\mathbf{h}$ is a vector, and$ h_0$ is a constant.

Gaussian systems can be simulated efficiently as their computational cost scales polynomially with the number of modes. This efficiency arises from their full characterization via covariance matrices and first moments, bypassing the need for exponentially large Hilbert space representations. These properties make Gaussian models indispensable for studying large-scale photonic quantum systems and serve as a baseline for incorporating non-Gaussian elements crucial for universal quantum computation.

\subsection{Non-Gaussian Models}

\begin{figure}[H]
\centering
\includegraphics[width=0.9\textwidth]{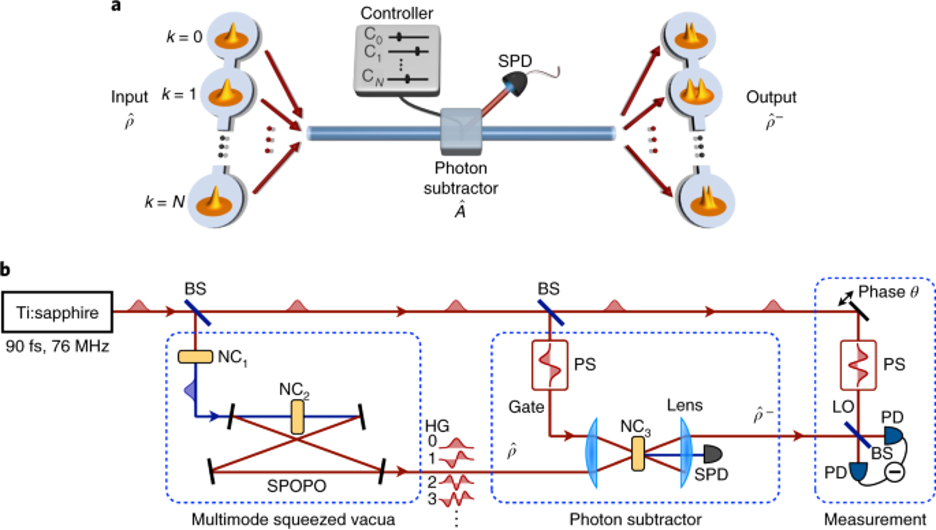}
\caption{Photon subtraction scheme used to generate non-Gaussian quantum states. (a) depicts the process where an input quantum state \(\hat{\rho}\) undergoes photon subtraction using a single-photon detector (SPD) and a controlled beam splitter. The output state \(\hat{\rho}^-\) exhibits enhanced non-Gaussian characteristics. (b) details the experimental setup, utilizing a Ti:sapphire laser (90 fs, 76 MHz) to generate multimode squeezed vacua inside a synchronously pumped optical parametric oscillator (SPOPO). Photon subtraction is performed via a beam splitter (BS), nonlinear crystal (NC), and SPD, followed by homodyne detection for quantum state measurement, adapted with permission from \citet{ra2020non}}
    \label{fig4}
\end{figure}

\begin{figure}[H]
\centering
\includegraphics[width=0.8\textwidth]{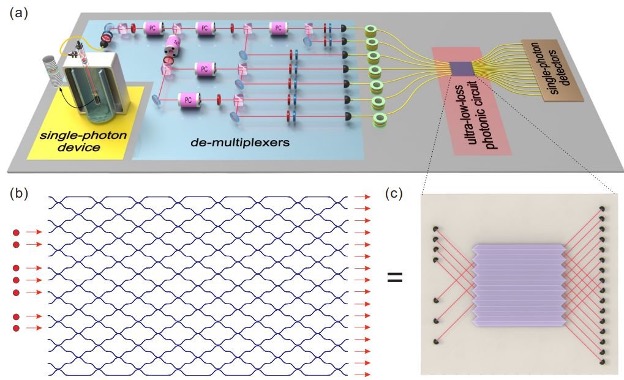}
\caption{This figure presents a lossy boson sampling experiment on an ultra-low-loss photonic chip. (a) illustrates the experimental setup, where a quantum-dot micropillar source at 4.2 K generates single photons, which are demultiplexed into seven spatial modes and time-compensated using single-mode fibers. These photons are injected into a 16 × 16 mode interferometric photonic circuit containing 113 beam splitters and 14 mirrors, ensuring high-fidelity quantum interference. (b) shows the theoretical model of the fully-connected interferometer with $>$ 99\% transmission efficiency, while (c) highlights its physical realization, crucial for scalable photonic quantum computing and boson sampling applications, adapted with permission from \citet{wang2018toward}}
    \label{fig5}
\end{figure}

Non-Gaussian states and operations in Figures \ref{fig4} and \ref{fig5} are fundamental to achieving universal quantum computation, as they enable computational capabilities that Gaussian systems alone cannot provide \cite{quesada2019simulating, genoni2007measure, ra2020non}. These states and operations break the mathematical simplicity of Gaussian frameworks and often require more complex representations.

\subsubsection{Non-Gaussian States and Operations} 
    \begin{enumerate}
        \item [a.]	Single-Photon States \cite{lvovsky2001quantum}: A single-photon state in mode j is represented in the Fock basis as:
    \end{enumerate}
    
\begin{equation}
    |1\rangle_j = \hat{a}_j^\dagger |0\rangle_j,
\end{equation}

where $|0\rangle_j$ is the vacuum state, and $\hat{a}_j^\dagger$ is the creation operator. The state $|1\rangle_j$ has a photon number of precisely one, making it non-Gaussian due to its discrete nature.

    \begin{enumerate}
        \item [b.]	Photon-Number-Resolving Detectors \cite{provaznik2020benchmarking}: These devices distinguish the photon number n in a given mode. The probability of detecting n photons is given by:
    \end{enumerate}

\begin{equation}
    P(n) = \langle n|\hat{\rho}|n\rangle,
\end{equation}

where $\hat{\rho}$ is the density matrix of the quantum state.

    \begin{enumerate}
        \item [c.]	Nonlinear Operations: Nonlinear operations, such as the cubic phase gate or photon subtraction, introduce non-Gaussianity. For instance:
    \end{enumerate}
    
    \begin{itemize}
        \item [i.] Cubic Phase Gate \cite{budinger2024all}: The cubic phase gate is defined by the unitary operator:
    \end{itemize}

\begin{equation}
    \hat{U}_{\text{cubic}} = \exp(i\gamma \hat{x}^3),
\end{equation}

where $\gamma$ is a parameter, and $\hat{x}$ is the position quadrature operator.

    \begin{itemize}
        \item [ii.] Photon Subtraction \cite{fan2018quantum}: Photon subtraction is modeled as the annihilation operator $\hat{a}$ acting on a state:
    \end{itemize}

\begin{equation}
    \hat{a}|n\rangle = \sqrt{n}|n-1\rangle.
\end{equation}

\subsubsection{Representations of Non-Gaussian States}

Non-Gaussian states cannot be described as Gaussian distributions in phase space, necessitating alternative frameworks:

    \begin{enumerate}
        \item [1.] Fock State Representation \cite{strocchi2021fock}: In the Fock basis, a general quantum state $|\psi\rangle$ is expressed as:
    \end{enumerate}

\begin{equation}
    |\psi\rangle = \sum_{n=0}^\infty c_n |n\rangle,
\end{equation}

where $c_n$ are the probability amplitudes. The computational complexity arises due to the exponential growth of the Hilbert space with the number of modes and photons.

    \begin{enumerate}
        \item [2.] Density Matrices \cite{mcweeny1960some}: For mixed states, the density matrix $\hat{\rho}$ is used:
    \end{enumerate}

\begin{equation}
    \hat{\rho} = \sum_{n,m} \rho_{nm} |n\rangle \langle m|.
\end{equation}

The size of $\hat{\rho}$ grows as $d^2$, where d is the dimension of the Hilbert space, making this representation computationally expensive for large systems.

    \begin{enumerate}
        \item [3.] Wigner Functions \cite{tatarskiui1983wigner,ohst2024symmetries}: Non-Gaussian states in phase space are described by Wigner functions $W(\mathbf{r})$, which may exhibit negativity, a signature of non-classicality:
    \end{enumerate}

\begin{equation}
    W(\mathbf{r}) = \frac{1}{(2\pi)^n} \int d^n\mathbf{s} \, \exp\left(-i \mathbf{r}^T \mathbf{\Omega} \mathbf{s}\right) \chi(\mathbf{s}),
\end{equation}

where$ \chi(\mathbf{s})$ is the characteristic function and $\mathbf{\Omega}$ is the symplectic form. The negativity of $W(\mathbf{r})$ complicates numerical simulations and distinguishes non-Gaussian states from Gaussian ones.

\subsubsection{Challenges in Simulating Non-Gaussian States}

Simulating non-Gaussian states presents significant challenges due to the exponential scaling of the Hilbert space dimension with the number of modes and photons. As the size of the system increases, the computational resources required to represent and evolve the quantum state grow exponentially, making exact simulations infeasible for large systems. Another challenge arises from the negativity in the Wigner function, a hallmark of non-classicality in non-Gaussian states. This negativity prevents the use of efficient classical sampling techniques that are effective for Gaussian systems, further complicating simulations. Additionally, non-Gaussian states are particularly sensitive to photon loss and noise, which degrade their unique quantum properties. These imperfections make it difficult to maintain and simulate the desired non-Gaussian features, posing a substantial hurdle for practical implementations.

\subsubsection{Computational Approaches}

To address these challenges, various computational techniques have been developed. Tensor network methods, such as matrix product states (MPS), provide a compact and efficient representation of weakly entangled quantum systems. These methods are particularly useful for simulating certain non-Gaussian states, as they can significantly reduce the computational complexity. Hybrid representations also play a crucial role in balancing accuracy and scalability. By combining Fock state representations for non-Gaussian components with covariance matrices for Gaussian parts, hybrid approaches capture the essential features of the system while mitigating computational demands.

Stochastic sampling methods, including Monte Carlo techniques, are frequently employed to approximate the behavior of non-Gaussian states in specific regimes. These methods allow for probabilistic exploration of the system’s state space, providing a practical alternative to exact simulations. Additionally, machine learning has emerged as a powerful tool for modeling and simulating non-Gaussian quantum states. Neural networks, in particular, excel at approximating high-dimensional functions and can be trained to predict the behavior of complex quantum systems with remarkable accuracy. These computational approaches collectively provide a pathway for advancing the simulation of non-Gaussian states despite their inherent challenges. Table \ref{tab1} demonstrates the various kinds of gates used by simulators.

\begin{table}[!h]
\centering
\caption{Gaussian and Non-Gaussian Gates in Quantum Photonic Simulators}
\label{tab1}
\begin{tabular}{>{\hspace{0pt}}m{0.158\linewidth}>{\hspace{0pt}}m{0.379\linewidth}>{\hspace{0pt}}m{0.404\linewidth}} 
\toprule
\textbf{Simulator} & \textbf{Gaussian Gates} & \textbf{Non-Gaussian Gates} \\ 
\hline
Strawberry Fields \cite{killoran2019strawberry} & Displacement, Squeezing, Phase shift, Beam splitter, Rotation & Kerr gate, Cubic phase gate, Photon subtraction, Photon addition \\ 
\hline
Ansys Lumerical \cite{ansysfdtd} & N/A (focused on classical device-level modeling) & N/A \\ 
\hline
Piquasso \cite{kolarovszki2024piquasso} & Displacement, Squeezing, Beam splitter, Phase shift & Kerr gate, Cubic phase gate, Photon subtraction, Photon addition \\ 
\hline
QuTiP \cite{QuTiP2015} & Any linear operator such as squeezing, displacement, rotation can be Gaussian & Any arbitrary custom operator for instance photon subtraction, and Fock-state manipulation\\ 
\hline
SimulaQron \cite{dahlberg2018simulaqron} & N/A (network-level simulator) & N/A \\ 
\hline
Perceval \cite{heurtel2023perceval} & Beam splitter, Phase shifter, Rotation, Displacement & Photon-number resolving gates, Post-selection gates \\ 
\hline
QuantumOptics.jl \cite{kramer2018quantumoptics} & Squeezing, Displacement, Beam splitter, Phase shift & Kerr gate, Cubic phase gate, Photon subtraction, Photon addition \\ 
\hline
Synopsys Photonic Solutions \cite{synopsysphotonic} & N/A (focused on classical photonic devices) & N/A \\
\bottomrule
\end{tabular}
\end{table}

\subsection{Continuous-Variable Frameworks}

The continuous-variable (CV) framework \cite{menicucci2006universal} is a bedrock of quantum photonics, offering a mathematical description of quantum systems based on quadrature operators $\hat{x}$ (position) and $\hat{p}$ (momentum). These operators are the quantum analogs of classical position and momentum, providing a natural framework for describing quantum states of light.

\subsubsection{Quadrature Operators and Phase Space}

The quadrature operators $\hat{x}$ and $\hat{p}$ are defined in terms of the annihilation ($\hat{a}$) and creation ($\hat{a}^\dagger$) operators as:

\begin{equation}
    \hat{x} = \frac{1}{\sqrt{2}} (\hat{a} + \hat{a}^\dagger), \quad \hat{p} = \frac{i}{\sqrt{2}} (\hat{a}^\dagger - \hat{a}).
\end{equation}

These operators satisfy the canonical commutation relation:
\begin{equation}
    [\hat{x}, \hat{p}] = i,
\end{equation}

where $\hbar$ = 1 in natural units. In this framework, quantum states are represented in phase space as functions of (x, p), with the Wigner function W(x, p) providing a quasi-probability distribution.

The annihilation operator $\hat{a}$ and its Hermitian conjugate $\hat{a}^\dagger$ act on the Fock basis states $|n\rangle$ as:

\begin{equation}
   \hat{a}|n\rangle = \sqrt{n} |n-1\rangle, \quad \hat{a}^\dagger |n\rangle = \sqrt{n+1} |n+1\rangle. 
\end{equation}

\subsubsection{Covariance Matrices}

Gaussian states in the CV framework are fully characterized by their first and second moments. The first moments form the displacement vector:

\begin{equation}
    \mathbf{d} = \langle \mathbf{R} \rangle, \quad \mathbf{R} = (\hat{x}_1, \hat{p}_1, \dots, \hat{x}_n, \hat{p}n)^T.
\end{equation}

The second moments are encoded in the covariance matrix $\boldsymbol{\sigma}$, defined as:

\begin{equation}
    \sigma{ij} = \frac{1}{2} \langle \hat{R}_i \hat{R}_j + \hat{R}_j \hat{R}_i \rangle - \langle \hat{R}_i \rangle \langle \hat{R}_j \rangle,
\end{equation}

where $\mathbf{R}$ represents the quadrature operator vector for an n-mode system.

\subsubsection{Symplectic Evolution of Covariance Matrices}

The evolution of a Gaussian state under a Gaussian operation is determined by the symplectic transformation of its covariance matrix. A Gaussian operation is represented by a symplectic matrix $\mathbf{S}$ that satisfies the relation:

\begin{equation}
    \mathbf{S} \boldsymbol{\Omega} \mathbf{S}^T = \boldsymbol{\Omega}, \quad \boldsymbol{\Omega} = \bigoplus_{i=1}^n \begin{pmatrix} 0 & 1 \\ -1 & 0 \end{pmatrix}.
\end{equation}

Under a Gaussian transformation, the covariance matrix evolves as:

\begin{equation}
    \boldsymbol{\sigma}{\prime} = \mathbf{S} \boldsymbol{\sigma} \mathbf{S}^T + \mathbf{N},
\end{equation}

where $\mathbf{N}$ represents additional noise introduced by the transformation. This evolution is particularly efficient to compute, scaling polynomially with the number of modes, making it ideal for simulating large Gaussian systems.

\subsubsection{Gaussian Boson Sampling}

\begin{figure}[H]
\centering
\includegraphics[width=1\textwidth]{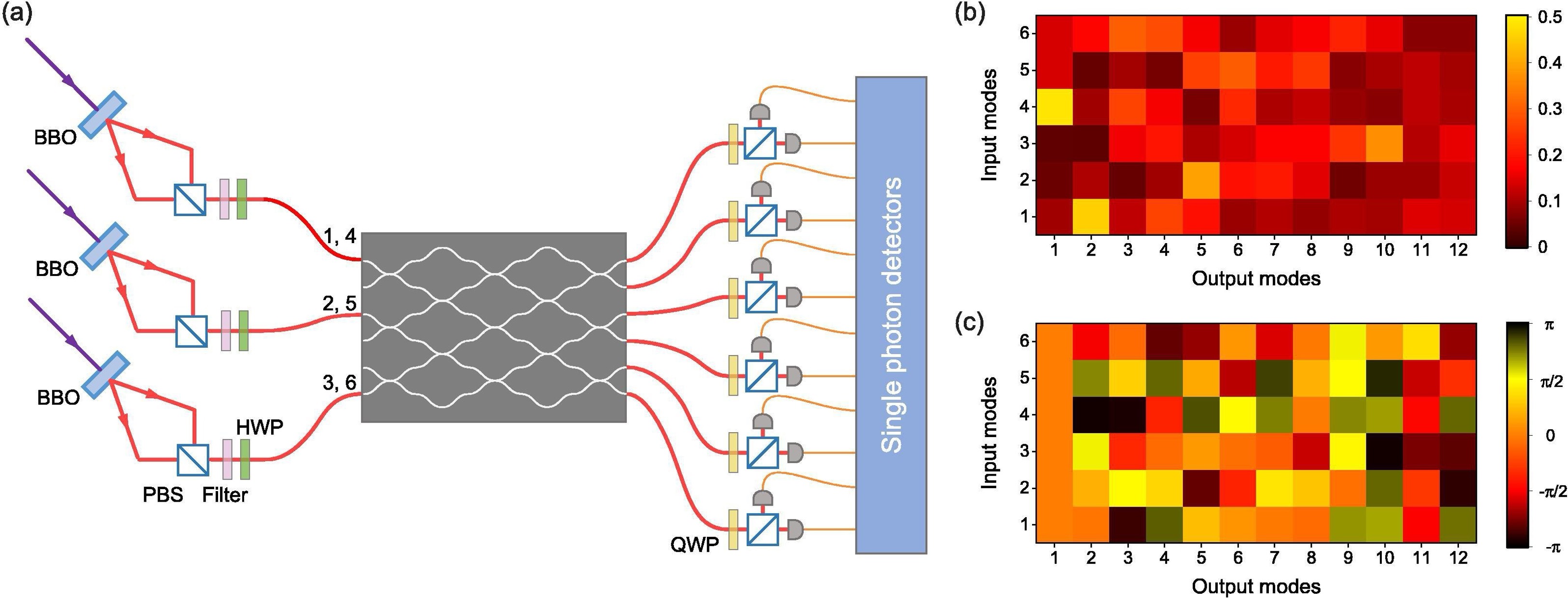}
\caption{Zhong illustrates a multiphoton quantum interference experiment in a bosonic network. (a) shows the experimental setup where entangled photon pairs are generated via spontaneous parametric down-conversion (SPDC) in BBO crystals. The photons are processed through a polarizing beam splitter (PBS), half-wave plate (HWP), and filters, then injected into a multi-mode interferometric network, where output photons are detected. (b) represents the probability distribution of photon correlations across input and output modes, while (c) depicts the corresponding phase information. This setup enables high-dimensional photonic quantum simulations and boson sampling experiments for quantum computational advantage, adapted with permission from \citet{zhong2019experimental}}
    \label{fig6}
\end{figure}

A critical application of the CV framework is Gaussian boson sampling \cite{hamilton2017gaussian,zhong2019experimental}, a computationally hard problem for classical systems but efficiently implemented using photonic quantum devices. In Gaussian boson sampling as carefully presented by Yan in Figure \ref{fig6}, the input state is a multimode Gaussian state characterized by its covariance matrix. The system undergoes a linear optical transformation (described by a symplectic matrix $\mathbf{S}$), followed by photon-number-resolving detection at the output. The probability of observing a specific photon-number distribution is determined by the Hafnian of a submatrix of the covariance matrix:

\begin{equation}
    P(n_1, n_2, \dots, n_m) \propto \text{Haf}(\boldsymbol{\sigma}\text{sub}),
\end{equation}

where $\boldsymbol{\sigma}\text{sub}$ is a submatrix of the covariance matrix, and $n_i$ represents the photon number in the $i-th$ mode. The computation of the Hafnian is classically hard, highlighting the computational advantage of quantum photonic systems.

The CV framework is particularly suited for implementing Gaussian boson sampling and modeling Gaussian quantum systems due to its mathematical tractability and scalability. It allows for efficient simulation of large-scale quantum systems using tools such as covariance matrices and phase-space representations, avoiding the exponential complexity associated with non-Gaussian quantum states.

\subsection{Noise and Decoherence}

Noise and decoherence are unavoidable phenomena in quantum photonic systems, arising from interactions with the environment, imperfections in components, and limitations of measurement devices. Accurate modeling of these effects is essential for assessing the robustness and performance of quantum circuits, as well as designing error-tolerant quantum systems.

\subsubsection{Gaussian Noise}

\begin{figure}[H]
\centering
\includegraphics[width=0.6\textwidth]{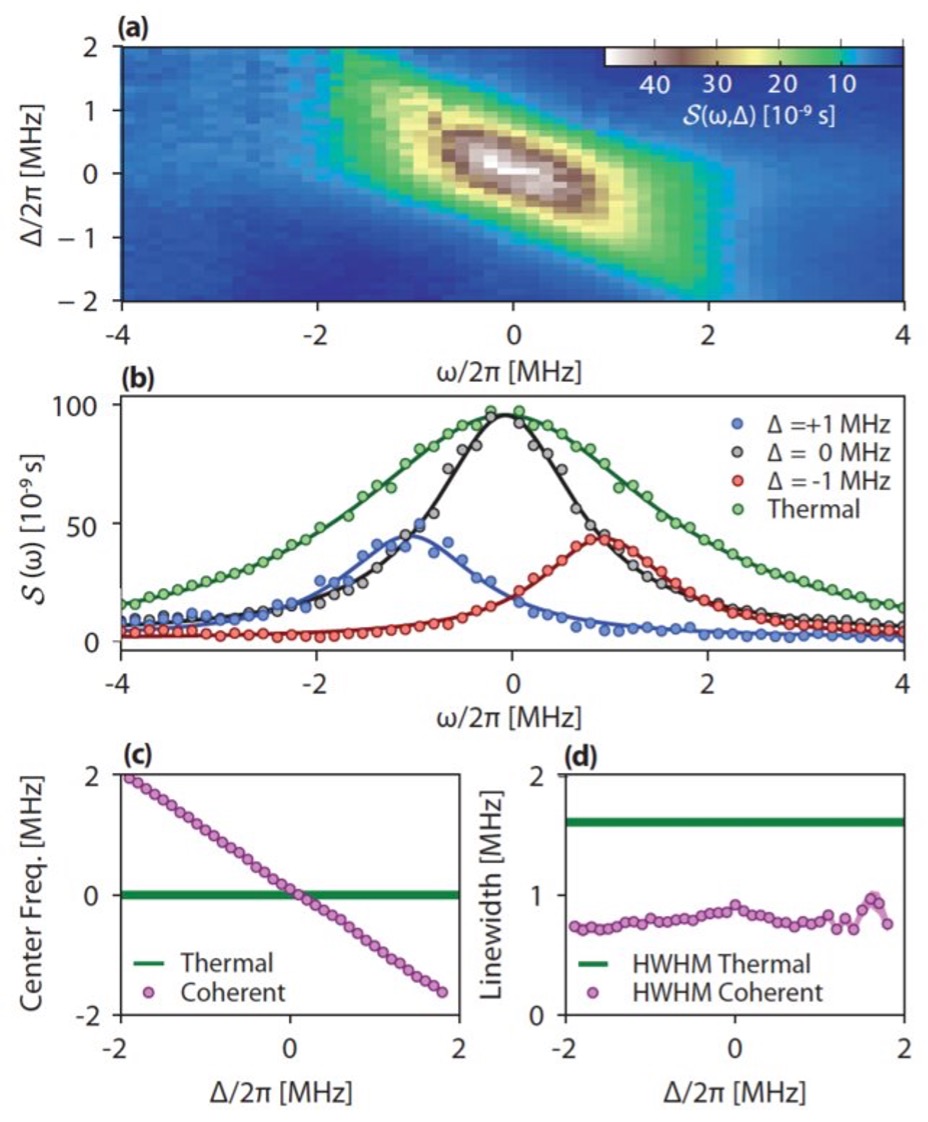}
\caption{Yan presents a spectral analysis of coherent and thermal states in a quantum system. (a) shows a heatmap of the spectral density \( S(\omega, \Delta) \), where frequency detuning (\(\Delta\)) and angular frequency (\(\omega\)) define the spectral response. (b) displays spectral distributions for different detunings (+1 MHz, 0 MHz, -1 MHz) and a thermal state highlighting line shape variations. (c) compares the center frequency shift of coherent (purple) and thermal (green) states. (d) analyzes the half-width at half-maximum (HWHM showing that thermal states exhibit broader linewidths than coherent states, indicating stronger decoherence effects, adapted with permission from \citet{yan2018distinguishing}}
    \label{fig7}
\end{figure}

Gaussian noise \cite{hall1994gaussian, yan2018distinguishing} describes processes such as attenuation and thermal effects as shown in Figure \ref{fig7}, are common in quantum photonic computing. These processes can be efficiently modeled within the continuous-variable framework by modifying the covariance matrix or the density operator.

For attenuation, the quantum system interacts with a lossy channel characterized by a transmissivity parameter $\eta$, where 0 $\leq$ $\eta$ $\leq 1$. In phase space, this modifies the covariance matrix $\boldsymbol{\sigma}$ as:

\begin{equation}
    \boldsymbol{\sigma}{\prime} = \eta \boldsymbol{\sigma} + (1 - \eta) \mathbf{I},
\end{equation}

where $\mathbf{I}$ is the identity matrix representing the vacuum noise introduced by the environment.

On the other hand, thermal noise introduces additional noise corresponding to a thermal state with mean photon number $\bar{n}$. In this case, the noise contribution is represented as:

\begin{equation}
    \mathbf{N} = (2\bar{n} + 1) \mathbf{I},
\end{equation}

where the covariance matrix evolves as:

\begin{equation}
    \boldsymbol{\sigma}{\prime} = \mathbf{S} \boldsymbol{\sigma} \mathbf{S}^T + \mathbf{N}.
\end{equation}

In the density operator formalism, thermal noise is modeled as a partial trace over a larger system or a convolution of the initial state with the thermal state’s density matrix:

\begin{equation}
    \hat{\rho}{\prime} = \int d^2\alpha \, P_\text{th}(\alpha) \hat{D}(\alpha) \hat{\rho} \hat{D}^\dagger(\alpha),
\end{equation}

where $P_\text{th}(\alpha)$ is the thermal state’s probability distribution and $\hat{D}(\alpha)$ is the displacement operator.

\subsubsection{Photon Loss}

\begin{figure}[H]
\centering
\includegraphics[width=0.7\textwidth]{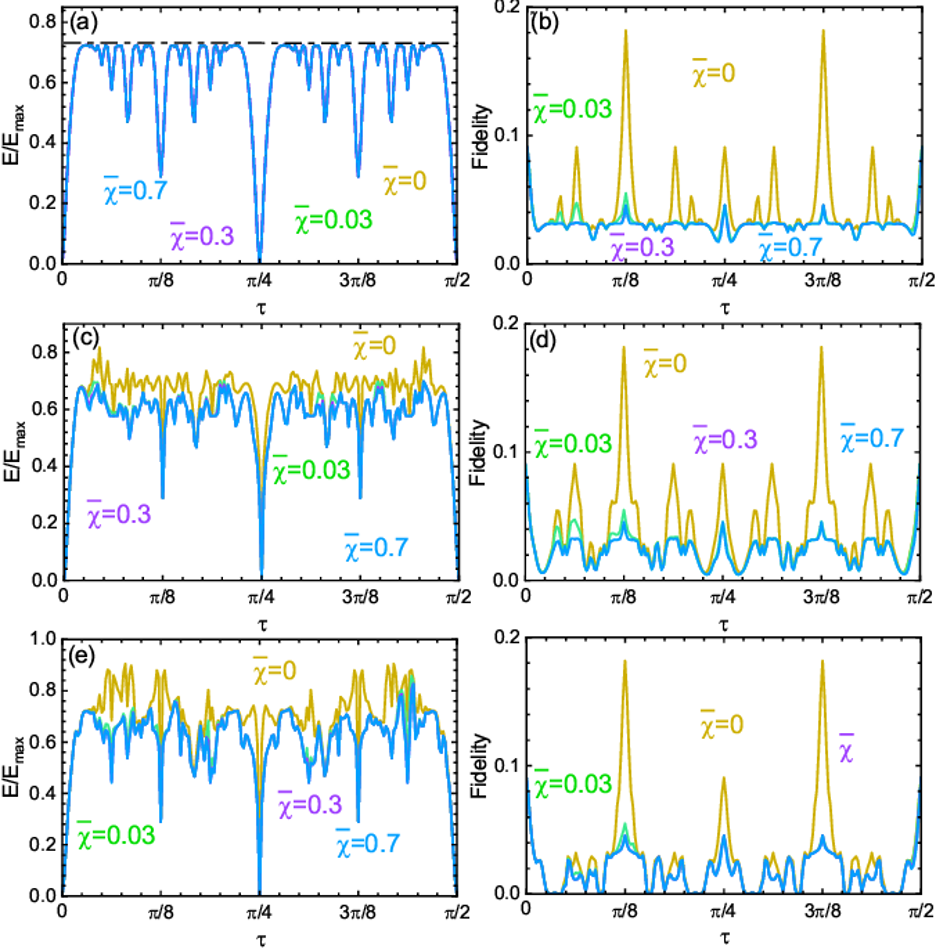}
\caption{Gao examines how photon loss affects entanglement and fidelity in a quantum system. (a, c, e) depict entanglement as a function of interaction time \(\tau\) for different photon measurement outcomes (\(n_d, n_c\)), showing that entanglement decreases with increasing loss. The dashed line represents the maximum achievable entanglement (b, d, f) illustrate fidelity degradation over time, indicating that higher photon loss where \(n_d = 70, n_c = 30\) leads to a more significant decline in quantum state preservation. This analysis highlights the impact of amplitude attenuation on quantum coherence and computational robustness, adapted with permission from \citet{gao2024photon}}
    \label{fig8}
\end{figure}

Photon loss as examined by Gao in Figure \ref{fig8} is a non-Gaussian noise process \cite{straka2014quantum,gao2024photon} that reduces the number of photons in a mode. It can be modeled using Kraus operators $\hat{K}_n$, which describe the action of a lossy channel with transmissivity $\eta$ on the Fock basis:

\begin{equation}
    \hat{K}n = \sqrt{\binom{n}{k} (1-\eta)^k \eta^{n-k}} |k\rangle\langle n|,
\end{equation}

where n and k are the photon numbers before and after loss, respectively. The density matrix after photon loss is given by:

\begin{equation}
    \hat{\rho}{\prime} = \sum{k=0}^n \hat{K}_k \hat{\rho} \hat{K}_k^\dagger.
\end{equation}

In stochastic simulations, photon loss is modeled as a probabilistic process where each photon is retained with probability $\eta$. The expected photon number $\langle \hat{n} \rangle$ after loss is scaled as:

\begin{equation}
    \langle \hat{n}{\prime} \rangle = \eta \langle \hat{n} \rangle.
\end{equation}

\subsubsection{Dark Counts}
\begin{figure}[H]
\centering
\includegraphics[width=0.8\textwidth]{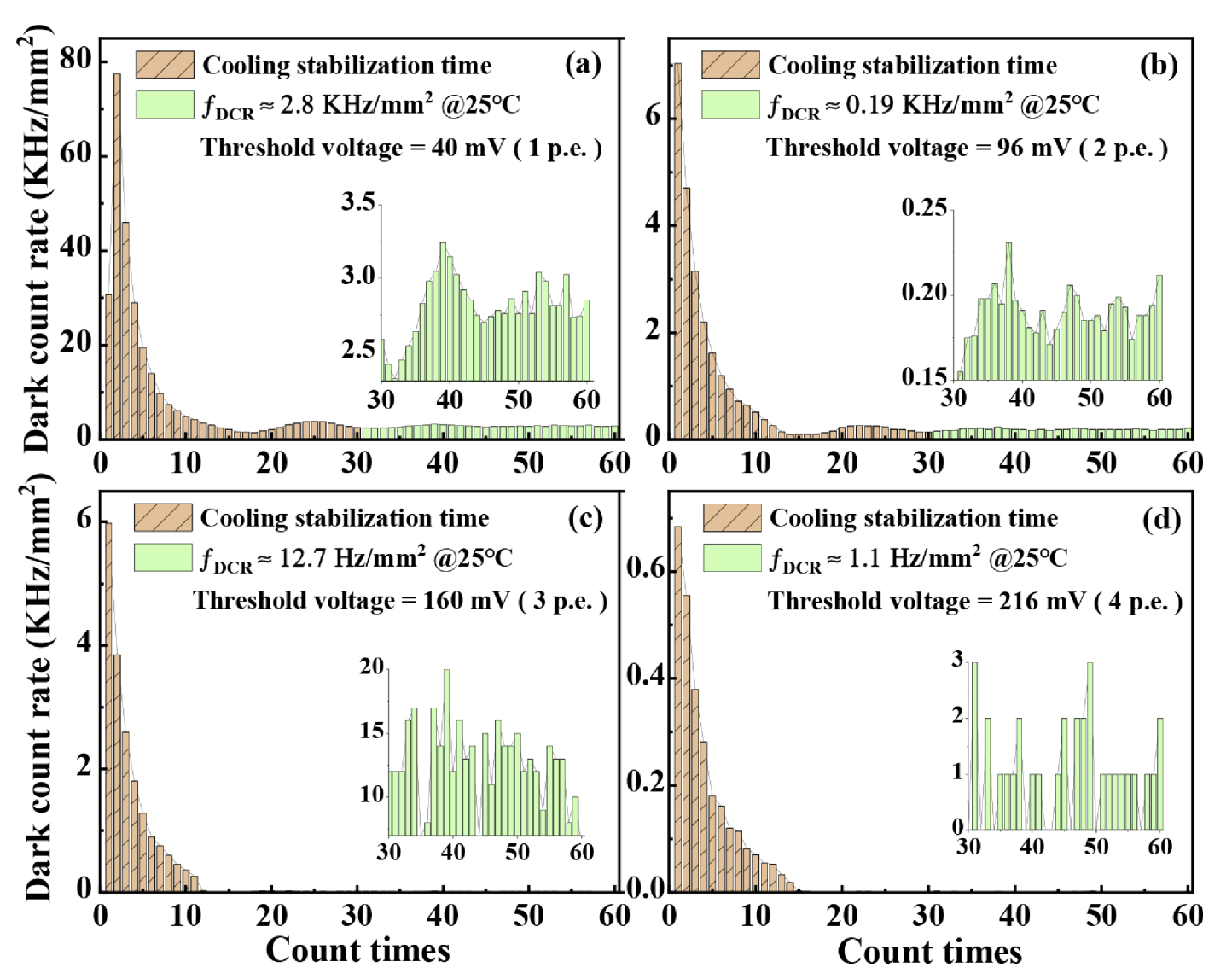}
\caption{Measured dark count rate (DCR of a silicon photomultiplier (SiPM) at different photoelectron (p.e.) threshold levels. (a-d) correspond to increasing threshold values: (a) 1 p.e. (40 mV), (b) 2 p.e. (96 mV), (c) 3 p.e. (160 mV), and (d) 4 p.e. (216 mV). The DCR decreases as the threshold increases, demonstrating that higher thresholds effectively suppress thermal noise and spurious dark counts. The brown-shaded regions represent the cooling stabilization period, ensuring stable detector performance. However, higher thresholds may reduce sensitivity, filtering out lower-amplitude photon signals in quantum detection applications, adapted with permission from  \citet{li2021single}.}
\label{fig9}
\end{figure}

Dark counts, as studied by \citet{li2021single} in Figure \ref{fig9} originate from imperfections in photon-number-resolving detectors \cite{kitaygorsky2005origin}, where spurious photon detections occur even in the absence of incident photons. These false counts primarily arise from thermal noise, afterpulsing effects, and electronic fluctuations within the detector. The occurrence of dark counts can be statistically modeled using a Poisson distribution:

\begin{equation}
    P_\text{dark}(k) = \frac{\lambda^k e^{-\lambda}}{k!},
\end{equation}

where \( \lambda \) represents the mean dark count rate, and \( k \) denotes the number of spurious detections.

In a realistic photon detection scenario, the total probability of measuring \( n \) detected photons is the convolution of the ideal photon distribution \( P_\text{ideal}(n) \) with the dark count probability:

\begin{equation}
    P_\text{measured}(n) = \sum_{k=0}^\infty P_\text{ideal}(n-k) P_\text{dark}(k).
\end{equation}

This correction accounts for false-positive photon detections due to background noise, providing a more accurate representation of the actual measured photon statistics.

\subsubsection{Inclusion of Noise in Simulations}

Incorporating noise into quantum photonic simulations is crucial for evaluating the performance of photonic circuits and developing error-tolerant quantum systems. By simulating noise \cite{vischi2024simulating} processes like Gaussian noise, photon loss, and dark counts, researchers can analyze their impact on the fidelity of quantum operations and identify strategies for mitigating their effects.

Error mitigation techniques \cite{ravi2022vaqem,cai2023quantum,bultrini2023unifying,ntone2024enhanced,endo2018practical,qin2022overview,strikis2021learning}, such as redundancy encoding and probabilistic error correction, are often combined with noise models to enhance the robustness of quantum circuits. These simulations enable a better understanding of the interplay between noise and circuit performance, providing insights into the design of practical quantum photonic devices.

\subsection{ Hybrid Quantum-Classical Systems}

Hybrid quantum-classical systems \cite{li2017hybrid} are an emerging paradigm that leverages the complementary strengths of quantum photonic circuits and classical computational resources. These systems integrate quantum processors for tasks that benefit from quantum parallelism and entanglement with classical optimization and machine learning algorithms, which excel in parameter optimization and large-scale data processing. This synergy enhances the applicability of photonic quantum simulators to tackle real-world challenges in fields such as sampling, optimization, and machine learning.

\subsubsection{Major Components of Hybrid Systems}
    \begin{enumerate}
        \item [a.] Variational Quantum Algorithms (VQAs) \cite{cerezo2021variational,perez2024variational}: 
        
        Variational quantum algorithms employ a classical optimizer to iteratively tune the parameters of a quantum circuit. The quantum circuit evaluates an objective function, while the classical optimizer updates the parameters to minimize or maximize the function. This approach is widely used for quantum chemistry, combinatorial optimization, and machine learning tasks.
        \item [b.] Tensor Networks \cite{yuan2021quantum}: 
        
        Tensor networks provide a compact representation of high-dimensional quantum states by exploiting the structure of entanglement. In hybrid systems, tensor networks are used to model and simulate large quantum systems efficiently, enabling scalable hybrid workflows that incorporate both quantum and classical components.
        \item [c.] Hardware Acceleration \cite{bertels2020quantum,beck2024integrating}: 
        
        The integration of classical hardware accelerators, such as GPUs and TPUs, with quantum photonic circuits significantly boosts computational performance. GPUs handle classical tasks like gradient computation and tensor manipulation, while quantum circuits execute quantum operations, ensuring efficient utilization of computational resources.
    \end{enumerate}

These systems exemplify the potential of hybrid approaches to bridge the gap between quantum and classical computing, expanding the scope and applicability of photonic quantum simulators.

\section{Survey of Existing Quantum Photonic Simulators}

The development of quantum photonic simulators has significantly advanced the design and exploration of photonic quantum systems. These tools aim to model the behavior of photons in circuits, enabling the design of quantum algorithms, hardware validation, and hybrid quantum-classical workflows. Several simulators have emerged as leading platforms, each with unique strengths, limitations, and areas of focus.

One of the most prominent simulators is Xanadu’s Strawberry Fields, a Python-based platform specifically designed for continuous-variable (CV) quantum computing. Strawberry Fields supports Gaussian and non-Gaussian states and integrates seamlessly with PennyLane, a quantum machine learning library. Its flexible Python API makes it accessible to researchers from various domains. Another important tool is Ansys lumerical, a commercial software widely used in the photonics industry for simulating the physical properties of photonic devices and circuits. While Ansys lumerical excels in device-level simulations, its quantum capabilities are limited, focusing primarily on classical photonic systems.

Simulators like SimulaQron provide infrastructure for simulating distributed quantum networks, which include photonic components. However, their primary focus lies in network-level operations rather than circuit-level modeling. Additionally, open-source tools such as QuTiP (Quantum Toolbox in Python) offer generic quantum simulation capabilities that can be adapted for photonic systems, though they are not specifically tailored for photonic quantum circuits.

\subsection{Photonic Simulators' Features and Capabilities}

The core features of quantum photonic simulators determine their applicability to specific tasks. For instance, Strawberry Fields stands out for its ability to model continuous-variable systems and perform Gaussian boson sampling. Its TensorFlow backend enables hybrid quantum-classical workflows, making it ideal for exploring variational quantum algorithms and machine learning applications. However, its scalability is limited when simulating large non-Gaussian systems due to the computational complexity of handling Fock states.

Ansys lumerical, on the other hand, focuses on the physical layer, offering advanced tools for waveguide design, beam splitter optimization, and modeling light propagation. While it does not natively support quantum computations, its ability to simulate the physical behavior of photonic components makes it a valuable tool for researchers designing quantum hardware. Tools like Ansys lumerical are often used in conjunction with quantum-focused simulators to bridge the gap between device-level and circuit-level modeling. In Table \ref{tab2} we provide other competitive photonic simulators that either utilize CV, DV, or in both hybrid states.

\begin{table}[!ht]
\centering
\caption{Comparison of Quantum Photonic Simulators by Features}
\label{tab2}
\begin{tabular}{>{\hspace{0pt}}m{0.102\linewidth}>{\hspace{0pt}}m{0.194\linewidth}>{\hspace{0pt}}m{0.183\linewidth}>{\hspace{0pt}}m{0.154\linewidth}>{\hspace{0pt}}m{0.135\linewidth}>{\hspace{0pt}}m{0.167\linewidth}} 
\toprule
\textbf{Simulator} & \textbf{Operators} & \textbf{Gates} & \textbf{States} & \textbf{Measurements} & \textbf{Decompositions} \\ 
\hline
Strawberry Fields \cite{killoran2019strawberry}  & Displacement, squeezing, phase shift & Gaussian and non-Gaussian gates such as cubic phase gate & Coherent, squeezed, thermal, Fock & Photon counting, homodyne, heterodyne & Symplectic, Takagi, Cholesky \\ 
\hline
Ansys Lumerical cite{ansysfdtd} & Classical wave equations (Maxwell-based) & N/A (focused on physical photonic devices) & Continuous classical optical modes & N/A (no quantum measurement tools) & N/A (not applicable for quantum systems) \\ 
\hline
Piquasso \cite{kolarovszki2024piquasso} & Displacement, squeezing, beam splitter & Gaussian and non-Gaussian gates for instance Kerr gate, cubic phase gate & Coherent, squeezed, thermal, Fock, and custom states & Photon counting, homodyne, heterodyne & Symplectic, Cholesky, user-defined custom decompositions \\ 
\hline
QuTiP \cite{QuTiP2015} & Arbitrary operators via dense or sparse matrices & Universal quantum gates (both DV and CV) & Any state represented in density matrix or ket form & Projective measurements & Matrix diagonalization, eigenvalue decomposition \\ 
\hline
SimulaQron \cite{dahlberg2018simulaqron} & Operators for distributed systems (e.g., entanglement swapping) & Network-level gates (e.g., Bell-state preparation) & Qubits and basic distributed states & Classical-quantum hybrid measurements & N/A (limited to network protocols) \\ 
\hline
Perceval \cite{heurtel2023perceval} & Beam splitters, phase shifters, displacement & Linear optical gates (interferometers) & CV and DV states, hybrid states & Photon-number resolving, homodyne & Singular value, polar decomposition for circuit analysis \\ 
\hline
QuantumOptics.jl \cite{kramer2018quantumoptics} & Arbitrary operators defined by users & Universal gates; supports CV and DV gates & Coherent, squeezed, Fock, and thermal states & Photon counting, homodyne, heterodyne & Eigen and Schmidt decompositions \\ 
\hline
Synopsys Photonic Solutions \cite{synopsysphotonic} & Classical light-matter interactions & N/A (focused on classical device-level gates) & Continuous classical optical fields & N/A (no quantum measurement tools) & N/A (not applicable for quantum simulations) \\
\bottomrule
\end{tabular}
\end{table}

Despite their utility, existing quantum photonic simulators face significant limitations. Scalability is one of the most pressing challenges, particularly for simulators handling non-Gaussian systems. Modeling large circuits with multiple non-Gaussian components, such as photon-number-resolving detectors or non-classical state generators, requires exponential computational resources. While tools like Strawberry Fields have made strides in incorporating non-Gaussian states, their scalability remains a bottleneck. Another challenge is the lack of seamless integration between photonic quantum simulators and physical device modeling tools. Researchers often need to use separate platforms for designing photonic hardware on Ansys lumerical and simulating quantum circuits using Strawberry Fields, resulting in inefficiencies and potential mismatches between theoretical designs and practical implementations.

Furthermore, the steep learning curve of some simulators can deter researchers and educators who lack advanced programming or mathematical expertise. While Python-based tools like Strawberry Fields and QuTiP offer accessible APIs, they still require a solid understanding of quantum mechanics and computational techniques. More intuitive, user-friendly interfaces are necessary to democratize access to these tools and enable broader adoption.

\subsection{Comparative Analysis}

A comprehensive evaluation of essential quantum photonic simulators in Table \ref{tab3} reveals their unique strengths and limitations, highlighting their applicability for specific tasks within quantum photonics. 

\begin{table}[!h]
\centering
\caption{Comparison of Quantum Photonic Simulators by Features and Popularity}
\label{tab3}
\begin{tabular}{>{\hspace{0pt}}m{0.125\linewidth}>{\hspace{0pt}}m{0.392\linewidth}>{\hspace{0pt}}m{0.285\linewidth}>{\hspace{0pt}}m{0.137\linewidth}} 
\toprule
\textbf{Simulator} & \textbf{Strengths} & \textbf{Limitations} & \textbf{Popularity (\% Engagement)} \\ 
\hline
Strawberry Fields \cite{killoran2019strawberry}  & Best for CV quantum computing; TensorFlow backend supports machine learning integration. & Limited scalability for large non-Gaussian systems. & 40\% \\ 
\hline
Ansys Lumerical \cite{ansysfdtd} & Excels in device-level simulations of optical components. & Lacks quantum computing-specific tools. & 30\% \\ 
\hline
Piquasso \cite{kolarovszki2024piquasso} & Supports Gaussian and non-Gaussian states and operations for CV and DV systems; flexible and user-friendly. & Still gaining traction; community resources are developing. & 10-15\% \\
\hline
QuTiP \cite{QuTiP2015} & Adaptable for photonic systems; supports general quantum simulations. & Lacks dedicated features for photonic circuits. & 25\% \\ 
\hline
SimulaQron \cite{dahlberg2018simulaqron} & Designed for distributed quantum networks. & Less suitable for detailed photonic circuit simulations. & 10\% \\ 
\hline
Perceval \cite{heurtel2023perceval} & Combines CV and DV approaches; hybrid system simulation and optimization. & Limited hardware modeling capabilities. & 20\% \\ 
\hline
QuantumOptics.jl \cite{kramer2018quantumoptics} & Versatile for simulating both CV and DV photonic systems. & Requires familiarity with Julia programming. & 10\% \\ 
\hline
Synopsys Photonic Solutions \cite{synopsysphotonic} & Excellent for device-level modeling of photonic integrated circuits and optical components. & Focused on classical photonic devices; lacks native quantum simulation tools. & 35\% \\ 

\bottomrule
\end{tabular}
\end{table}

\subsection{Emerging Trends}

Recent advancements in simulation techniques are addressing some of these challenges. Tensor network methods, for instance, have emerged as a powerful approach for simulating large photonic systems. By exploiting the structure of entanglement in quantum states, tensor networks can efficiently represent and simulate systems that would otherwise be computationally intractable. Strawberry Fields has incorporated some aspects of tensor network techniques, though their full potential remains underutilized in most simulators.

GPU acceleration is another promising development. By parallelizing computational tasks, GPUs significantly enhance the performance of quantum photonic simulations. This approach is particularly effective for simulating Gaussian systems and hybrid quantum-classical workflows. However, integrating GPU acceleration with non-Gaussian simulations remains a technical challenge.

Machine learning techniques are also making their way into quantum photonic simulations. These methods are being used to optimize circuit designs, identify noise-resilient configurations, and even simulate complex quantum states. Tools like Strawberry Fields, with their TensorFlow backend, are well-positioned to leverage these advancements, but further development is needed to fully integrate machine learning capabilities into photonic simulators.

\section{Advancements in Simulation Techniques}
\subsection{Tensor Networks for Scalability in SoCs and HPC}

One of the most significant breakthroughs in quantum photonic simulation is the application of tensor networks to manage the exponential growth of computational resources required for large-scale quantum systems \cite{masot2024stabilizer}. Tensor networks allow for a compact representation of quantum states, particularly useful in simulating Gaussian and weakly entangled non-Gaussian states.

\begin{figure}[H]
\centering
\includegraphics[width=0.5\textwidth]{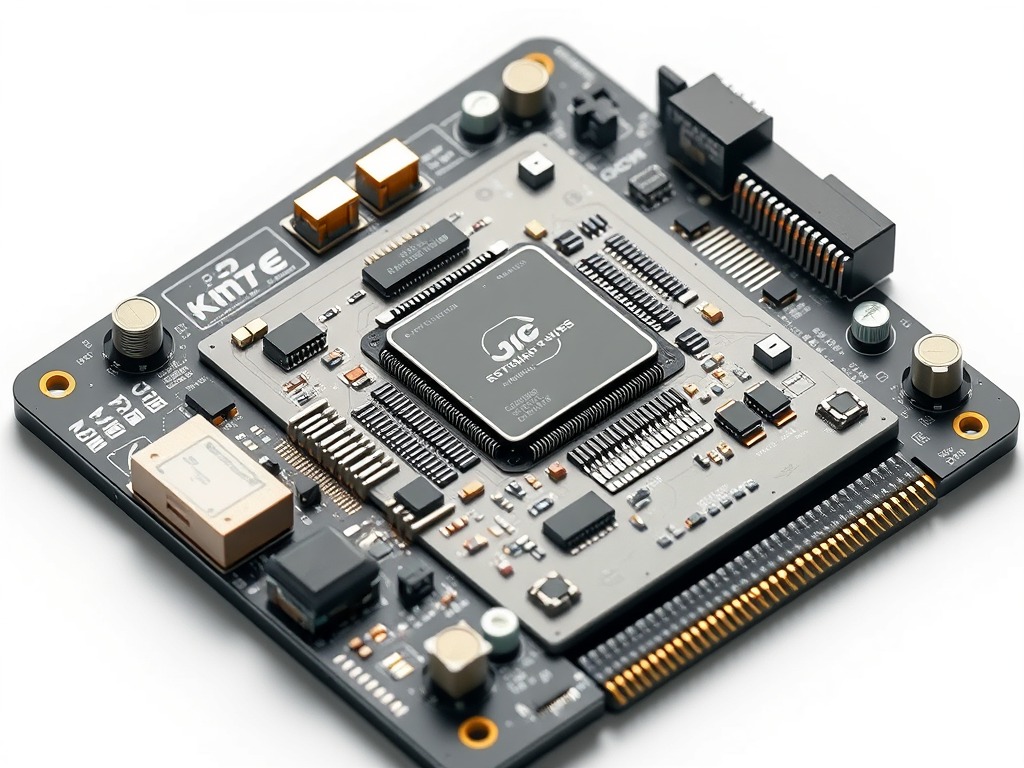}
\caption{This is an illustration of AI hardware development board, which most likely features an FPGA (Field-Programmable Gate Array) or ASIC (Application-Specific Integrated Circuit). It presents a promising platform for running a photonic quantum simulator on an SoC. With its high-speed parallel processing capabilities, low-latency memory interfaces, and customizable logic gates, this system can efficiently model quantum photonic circuits, handle tensor network calculations, and simulate Gaussian and non-Gaussian photonic states in real time. By leveraging neuromorphic and photonic co-processors, this architecture can support hybrid quantum-classical algorithms, accelerating tasks such as quantum state tomography, boson sampling, and error correction modeling essential for scalable photonic quantum computing research, adapted from \citet{Prabhakar2025}.}
\label{fig10}
\end{figure}

For SoCs as represented in Figure \ref{fig10}, tensor network optimizations are often constrained due to limited memory bandwidth and computational power, making them more suitable for small-scale quantum simulations or edge computing applications. SoCs such as Apple M2 Ultra and Google Edge TPU can leverage tensor-based AI-driven models but struggle with large-scale quantum photonic simulations.

In contrast, HPC architectures, including NVIDIA A100 GPUs and Google Quantum AI clusters, enable tensor network-based simulations at large scales by utilizing high-bandwidth memory and parallel processing capabilities. The bond dimensions in matrix product states (MPS) and Gaussian tensor networks can be significantly increased in HPC environments, providing higher accuracy in photonic circuit simulations.

\subsection{SoC and HPC Accelerated Parallel Computing}

The integration of parallel computing techniques in quantum photonic simulations has been significantly enhanced by both SoCs and HPC systems. While GPUs have historically driven acceleration, AI-optimized SoCs and TPUs are emerging as viable alternatives for photonic simulations, particularly in machine learning-assisted photonic circuit optimization \cite{faj2023quantum,willsch2022gpu}.

High-Performance Computing (HPC) GPUs, such as the NVIDIA A100, provide massive parallelism and enable efficient computation of matrix multiplications, eigenvalue decompositions, and tensor contractions, making them highly suitable for large-scale photonic simulations. These GPUs leverage their high memory bandwidth and optimized tensor processing capabilities to accelerate quantum photonic simulations that involve complex entanglement structures and large-scale non-Gaussian states.

On the other hand, AI-optimized SoCs, including NVIDIA Jetson and Google Edge TPU, are particularly effective for AI-assisted photonic circuit optimization. These systems excel in tasks such as quantum state classification, circuit parameter tuning, and real-time adaptive learning. However, they lack the necessary memory bandwidth and computational scalability required for full-scale quantum state evolution, limiting their role to small- to medium-scale simulations or edge computing applications.

Additionally, FPGA-accelerated SoCs, such as the Xilinx Versal AI Core, provide custom gate-level acceleration for photonic circuits. These systems are highly flexible, allowing researchers to implement specialized quantum algorithms and customized tensor network operations. However, they require specialized programming expertise and low-level hardware optimization, making them less accessible for general quantum photonic researchers compared to HPC-based solutions.

Beyond GPUs and TPU as illustrated in Tables \ref{tab4} and \ref{tab5}, recent advancements in NVIDIA cuQuantum SDK have extended tensor network optimization to HPC clusters, achieving up to 10x acceleration in large-scale quantum simulations \cite{bayraktar2023cuquantum,pan2024efficient,alevras2016gpu,hauru2021simulation,tornow2024quantum}.

\subsection{Machine Learning in SoC and HPC-Based Photonic Simulations}

Machine learning (ML) \cite{mahesh2020machine,zhou2021machine,alpaydin2021machine,jordan2015machine,fan2022optimizing,paler2023machine} has emerged as a powerful tool for optimizing photonic quantum circuits and noise modeling. ML techniques are being integrated into both SoCs and HPC systems, offering different advantages based on computational scale.

AI-optimized SoCs, such as Google Edge TPU, are well-suited for low-power, real-time machine learning tasks in photonic quantum computing. These chips are particularly effective for quantum state characterization and circuit optimization, where AI-driven approaches can enhance parameter tuning and adaptive learning. Their efficiency in handling real-time data processing makes them valuable for on-device quantum photonic applications \cite{gao2018experimental, vintskevich2023classification}, particularly in edge computing environments where power efficiency is a critical factor.

In contrast, HPC architectures, including NVIDIA A100 GPUs and Quantum AI Clusters, provide the necessary computational power to support deep learning models for advanced quantum photonic applications. These systems facilitate quantum state tomography, error mitigation strategies, and variational quantum algorithms (VQAs), enabling large-scale quantum simulations with high accuracy. HPC-based solutions are essential for tackling high-dimensional quantum systems that require extensive parallel processing, making them the preferred choice for researchers working on large-scale photonic quantum computing frameworks.

\subsection{Noise Modeling and Error Mitigation in SoC and HPC}

Both SoCs and HPC architectures play a role in noise modeling and error mitigation for quantum photonic simulators. The efficiency of these platforms in simulating photon loss, dark counts, and decoherence varies based on their hardware capabilities.

SoCs, such as Apple M2 Ultra, Google Edge TPU, and NVIDIA Jetson, offer a practical solution for low-latency, AI-driven noise prediction models in quantum photonic simulations. These systems are particularly useful in real-time applications where fast decision-making is required to adjust for small perturbations in quantum systems. However, their capabilities are limited when it comes to handling full quantum channel noise modeling, primarily due to their restricted memory bandwidth and computational constraints. As a result, while SoCs are beneficial for edge-based quantum noise estimation and adaptive correction, they struggle with complex noise propagation in large-scale photonic circuits.

In contrast, HPC architectures, such as Google Quantum AI clusters and NVIDIA A100 GPUs, provide the computational resources necessary for high-fidelity noise modeling. These systems utilize Kraus operators and tensor-based stochastic sampling techniques to accurately simulate decoherence, photon loss, and dark counts in quantum photonic circuits. Furthermore, HPC platforms are far more effective in simulating large-scale, fault-tolerant quantum architectures, making them indispensable for advanced research in photonic quantum error correction and large-scale quantum network simulations. By leveraging extensive parallel processing and high-memory bandwidth, HPC-based noise modeling enables precise evaluation of quantum system performance under realistic operational conditions.

\vfil

\subsection{Comparison of SoCs and HPC for Quantum Photonic Simulations}

To illustrate the differences in computational efficiency, we compare SoCs and HPC architectures based on their performance in quantum photonic simulations.

\begin{longtable}{>{\raggedright\arraybackslash}m{0.18\linewidth} 
                  >{\centering\arraybackslash}m{0.22\linewidth} 
                  >{\centering\arraybackslash}m{0.15\linewidth} 
                  >{\centering\arraybackslash}m{0.14\linewidth} 
                  >{\raggedright\arraybackslash}m{0.25\linewidth}}
\caption{Comparison of SoCs and HPC for Quantum Photonic Simulations\label{tab4}}\\ 
\toprule
\textbf{Hardware Type} & \textbf{Quantum Suitability} & \textbf{Processing Power (TFLOPS)} & \textbf{Memory Bandwidth (GB/s)} & \textbf{Parallel Processing Capabilities} \\ 
\midrule

NVIDIA Jetson AGX Orin \cite{NVIDIAJetsonOrin} & AI-driven quantum circuit optimization & 200 (AI/ML) & 204 & Optimized for AI, but lacks dedicated quantum evolution hardware \\ 
\hline
Apple M2 Ultra \cite{AppleM2Ultra2023} & Small-scale photonic simulations using OpenCL & 22 (GPU Compute) & 800 & Moderate parallelism, but lacks dedicated quantum acceleration \\ 
\hline
Google Edge TPU \cite{GoogleTPU2024} & AI-assisted quantum noise mitigation & 8 (Edge AI) & 32 & TPU optimized for AI applications, limited support for quantum workloads \\ 
\hline
Xilinx Versal AI Core \cite{XilinxVCK190} & Custom quantum photonic gate simulations & Custom Logic & Custom & Optimized for custom quantum gate acceleration but not large-scale quantum evolution \\ 
\hline
NVIDIA A100 \cite{NVIDIAA100} & Large-scale tensor network and quantum circuit calculations & 312 & 1,555 & High parallelism, well-suited for quantum tensor simulations \\ 
\hline
Google Quantum AI (HPC Cluster) \cite{GoogleQuantumAI} & Hybrid quantum-classical photonic simulations & 500+ (Cloud Distributed) & Variable & Cloud-based high-performance computing for quantum photonic workflows \\ 
\hline
Oak Ridge Summit (HPC Cluster) \cite{OLCF_HPC_Clusters} & Fault-tolerant quantum computing research & 1,000+ (HPC System) & 2,000+ & Extreme parallelism, essential for large-scale quantum error correction simulations \\
\bottomrule
\end{longtable}

\subsection{Real-Time Visualization and User Interfaces}

Visualization tools are indispensable for understanding and designing photonic quantum systems. Advancements in real-time visualization techniques have enabled researchers to observe photon dynamics, circuit configurations, and quantum state evolution during simulations.

Modern photonic simulators incorporate 2D and 3D visualization features to depict waveguides, beam splitters, and photon trajectories. Tools like Plotly \cite{sievert2020interactive}, VTK \cite{schroeder2000visualizing}, and Matplotlib \cite{tosi2009matplotlib} are often used to render these visualizations, providing intuitive representations of complex quantum phenomena. For instance, Wigner function plots are commonly used to visualize quantum states in phase space, while heatmaps can show photon density distributions in circuits.

Interactive user interfaces further enhance the usability of photonic simulators, allowing researchers to design circuits, adjust parameters, and view results in real time. SwiftUI and similar frameworks have been proposed for building intuitive front-end applications that cater to both researchers and educators. By combining visualization with interactivity, these advancements make quantum photonic simulations more accessible and comprehensible.

\subsection{Integration with Hybrid Ecosystems}

As photonic quantum computing progresses, the integration of simulators with hybrid quantum-classical ecosystems has become a priority. Tools like Xanadu’s Strawberry Fields have pioneered this approach, enabling seamless connectivity with machine learning libraries such as TensorFlow \cite{abadi2016tensorflow} and PyTorch \cite{imambi2021pytorch,ketkar2021introduction}.

Hybrid ecosystems facilitate workflows where classical algorithms preprocess data, quantum photonic circuits perform specialized computations, and classical postprocessing interprets the results. This paradigm is particularly relevant for variational quantum algorithms \cite{lubasch2020variational,benedetti2021hardware} and machine learning applications, where quantum and classical resources complement each other.

Interoperability with physical design tools, such as Ansys lumerical, has also become a focus area. By linking circuit-level simulations with device-level modeling, researchers can design and validate photonic circuits that are ready for experimental implementation. This integration streamlines the development pipeline and accelerates the transition from theoretical models to practical applications.

While significant progress has been made, challenges remain in scaling simulations to larger and more complex photonic systems. The development of hybrid tensor network-GPU frameworks, adaptive noise modeling algorithms, and ML-driven optimization tools represents promising avenues for future research.

Open-source initiatives and collaborative platforms will play a vital role in advancing simulation techniques. By democratizing access to cutting-edge tools, the quantum photonics community can foster innovation and accelerate progress. Additionally, efforts to standardize interfaces and data formats will enhance interoperability between simulators and other software ecosystems.

\begin{longtable}{>{\raggedright\arraybackslash}m{0.12\linewidth} 
                  >{\centering\arraybackslash}m{0.10\linewidth} 
                  >{\centering\arraybackslash}m{0.12\linewidth} 
                  >{\centering\arraybackslash}m{0.10\linewidth} 
                  >{\centering\arraybackslash}m{0.12\linewidth} 
                  >{\centering\arraybackslash}m{0.12\linewidth} 
                  >{\centering\arraybackslash}m{0.12\linewidth} 
                  >{\centering\arraybackslash}m{0.12\linewidth}}
\caption{Comparison of Quantum Photonic Simulators with Advanced Simulation Techniques} \label{tab5} \\ 
\toprule

\textbf{Simulator} & \textbf{Quantum Model} & \textbf{GPU Acceleration} & \textbf{TPU Support} & \textbf{Tensor Networks} & \textbf{Machine Learning} & \textbf{Noise Modeling} & \textbf{Hybrid Integration} \\
Strawberry Fields \cite{killoran2019strawberry}  & CV & \checkmark\ CUDA, TensorFlow & \ding{55} No TPU Support & \checkmark\ Gaussian Tensor Networks & \checkmark\ TensorFlow-based VQAs & \checkmark\ Kraus Operator Noise & \checkmark\ Integrated with PennyLane \\ 
\hline
Ansys Lumerical \cite{ansysfdtd} & Device-Level & \checkmark\ Hardware-Accelerated FDTD & \ding{55} No TPU Support & \ding{55} Not Tensor-Based & \ding{55} No ML Support & \checkmark\ Waveguide Roughness, Fabrication Noise & \ding{55} No Hybrid Support \\ 
\hline
Piquasso \cite{kolarovszki2024piquasso} & DV \& CV & \checkmark\ CUDA Support & \ding{55} No TPU Support & \checkmark\ MPS-Based Tensor Methods & \checkmark\ ML-Assisted Circuit Optimization & \checkmark\ Stochastic Noise Modeling & \checkmark\ Classical-Quantum Workflows \\ 
\hline
QuTiP \cite{QuTiP2015} & DV & \checkmark\ Limited GPU Support & \ding{55} No TPU Support & \ding{55} No Tensor Networks & \checkmark\ Neural Net Tomography & \checkmark\ Quantum Channel Noise & \ding{55} No Hybrid Support \\ 
\hline
SimulaQron \cite{dahlberg2018simulaqron} & Quantum Networks & \ding{55} No GPU Support & \ding{55} No TPU Support & \ding{55} Not Tensor-Based & \ding{55} No ML Support & \checkmark\ Network Channel Noise & \checkmark\ Networked Quantum-Classical Simulations \\ 
\hline
Perceval \cite{heurtel2023perceval} & CV & \checkmark\ GPU-Enabled & \ding{55} No TPU Support & \checkmark\ MPS and Gaussian Tensor Methods & \checkmark\ Variational Hybrid ML & \checkmark\ Adaptive Noise Filtering & \checkmark\ Classical-Photonic Simulation \\ 
\hline
Quantum
Optics.jl \cite{kramer2018quantumoptics}& Quantum Optics & \checkmark\ Julia GPU Libraries & \checkmark\ Experimental TPU Support & \ding{55} No Tensor Networks & \checkmark\ ML-Assisted Quantum State Classification & \checkmark\ Decoherence Modeling & \checkmark\ Optical Simulations with Classical Methods \\ 
\hline
Synopsys Photonic 
Solutions \cite{synopsysphotonic} & Device-Level & \checkmark\ Hardware-Accelerated FDTD & \ding{55} No TPU Support & \ding{55} No Tensor Networks & \ding{55} No ML Support & \checkmark\ Device-Level Noise Models & \checkmark\ Integrated with Classical Photonic Tools \\ 
\hline
\end{longtable}

\section{Challenges and Open Problems}

One of the most pressing challenges in quantum photonic simulation is scalability. Photonic systems, particularly those incorporating non-Gaussian elements, grow exponentially in complexity as the number of modes or photons increases. Gaussian simulations, while efficient, are limited in their ability to represent universal quantum computation, requiring the inclusion of non-Gaussian states such as photon subtraction or single-photon sources. Non-Gaussian simulations often rely on Fock state representations or density matrices, both of which demand exponential computational resources. This scalability issue is exacerbated when simulating hybrid systems that combine continuous-variable (CV) and discrete-variable (DV) elements.

Tensor networks and other advanced representations have been proposed to mitigate these issues, but they remain constrained by the entanglement and non-Gaussianity of the system. Bond dimensions in tensor networks, which govern the accuracy of the representation, must grow rapidly to capture complex quantum states, limiting their effectiveness for highly entangled systems. Developing adaptive tensor networks and hybrid approaches that dynamically adjust computational resources is an open area of research that could significantly enhance scalability.

\subsection{Noise and Decoherence Modeling}

Realistic quantum photonic systems are inherently noisy, with imperfections such as photon loss, phase errors, and detector inefficiencies degrading the fidelity of quantum operations. Accurately modeling these noise sources is critical for designing robust quantum circuits and evaluating their performance. Gaussian noise, such as thermal fluctuations and attenuation, can be efficiently modeled within the covariance matrix framework. However, non-Gaussian noise, such as photon loss and dark counts, requires more complex techniques such as Kraus operators or stochastic sampling.

One open problem in this area is the development of comprehensive noise models that capture the full spectrum of real-world imperfections, including fabrication errors in integrated photonic devices. These models must account for variability in waveguide geometries, phase-matching conditions, and coupling efficiencies. Incorporating these imperfections into simulations without sacrificing computational efficiency remains a major challenge \cite{brand2024markovian}.

Error mitigation techniques, such as redundancy encoding and probabilistic error cancellation, have shown promise in counteracting noise. However, their integration into photonic simulators is still in its infancy. Research into scalable error correction schemes and fault-tolerant architectures tailored to photonic quantum systems is critical for advancing practical implementations.

\subsection{Integration with Experimental Platforms}

The gap between theoretical simulations and experimental implementations poses another significant challenge. While simulators provide valuable insights into the design and behavior of photonic quantum systems, their predictions often fail to account for real-world complexities. Bridging this gap requires tight integration between simulation tools and experimental platforms.

One of the important challenges in this integration is the lack of interoperability between software platforms \cite{alfonso2024towards}. For instance, tools like Ansys lumerical and Ansys excel at modeling the physical properties of photonic components but lack quantum capabilities, while simulators like Strawberry Fields focus on circuit-level modeling with limited connections to physical design tools. Developing unified frameworks that combine device-level and circuit-level simulations would streamline the research and development pipeline.

Additionally, validating simulation results against experimental data is often hampered by the difficulty of measuring quantum states. Experimental uncertainties, noise, and limited detector capabilities make it challenging to obtain accurate benchmarks. Improved state tomography techniques and better calibration methods are needed to enhance the fidelity of experimental validation.

\subsection{Hybrid Quantum-Classical Workflows}

Hybrid quantum-classical systems are increasingly viewed as a practical approach to leveraging the strengths of both paradigms \cite{lubasch2018tensor}. In photonic quantum computing, hybrid workflows combine classical preprocessing and postprocessing with quantum operations performed by photonic circuits. While these workflows offer significant advantages, their simulation presents unique challenges.

One challenge lies in integrating classical algorithms with quantum photonic simulators \cite{shu2024general}. Classical components, such as machine learning models or optimization routines, must interact seamlessly with quantum simulations, often requiring specialized interfaces or middleware. Additionally, the computational cost of simulating hybrid systems can be prohibitive, as it combines the complexity of both quantum and classical computations.

Another open problem is the development of frameworks that support variational quantum algorithms (VQAs) and other hybrid approaches \cite{chen2024novel}. These frameworks must allow for dynamic adjustments of circuit parameters based on feedback from classical optimizers, which can introduce latency and synchronization issues in simulations. Tools like Strawberry Fields have made strides in this area, but more versatile and scalable solutions are needed.

\subsection{User Accessibility and Usability}

Quantum photonic simulators are powerful tools, but their complexity often limits their accessibility to a narrow group of experts. Researchers, educators, and students without extensive programming or quantum mechanics backgrounds may struggle to use these tools effectively. This lack of accessibility impedes the broader adoption of photonic simulation technologies and restricts the flow of innovation.

Developing intuitive user interfaces is a critical step toward addressing this challenge. Platforms like SwiftUI offer opportunities to create interactive, visually appealing interfaces that simplify circuit design and parameter configuration. These interfaces can lower the entry barrier for new users while still providing advanced features for experienced researchers.

Documentation, tutorials, and instance workflows are equally important for enhancing accessibility. Open-source initiatives have played a significant role in this regard, fostering collaborative communities where users can share knowledge and contribute to tool development. Expanding these efforts to include comprehensive educational resources will be prime to democratizing access to quantum photonic simulation.

\subsection{Advancements in Visualization}

Visualization is a foundation of understanding quantum photonic systems, yet existing tools often fall short of providing intuitive, real-time representations. Simulating quantum states, photon trajectories, and circuit layouts can produce vast amounts of data that are challenging to interpret without effective visualization techniques.

The development of advanced visualization tools that support 2D and 3D representations of circuits, waveguides, and quantum states is an open problem. For instance, real-time heatmaps showing photon density distributions or Wigner function plots visualizing quantum state properties can provide valuable insights. Integrating these features into photonic simulators without introducing significant computational overhead is a technical challenge that requires innovative solutions.

Additionally, visualization tools must cater to diverse user needs, ranging from researchers conducting detailed analyses to educators demonstrating basic quantum concepts. Developing modular visualization frameworks that adapt to different use cases is a promising approach to addressing this challenge.

\subsection{Open-Source Collaboration and Standardization}

The quantum photonics community has benefited greatly from open-source initiatives, but there is still room for improvement in collaboration and standardization. The lack of standardized interfaces and data formats across simulators creates inefficiencies and hinders interoperability. Establishing common protocols for data exchange, circuit descriptions, and simulation outputs would streamline research workflows and facilitate tool integration.

Open-source platforms have the potential to drive innovation by enabling researchers to build on existing tools rather than starting from scratch. However, maintaining and updating these platforms requires significant resources, including funding and community support. Encouraging institutional and industry participation in open-source projects is essential for sustaining their development.

\subsection{Ethical and Practical Implications}

As quantum photonic technologies advance, ethical and practical considerations are becoming increasingly important. The potential for quantum systems to disrupt fields like cryptography raises concerns about their misuse. Simulators, as tools for designing and testing quantum systems, must include safeguards to prevent malicious applications.

Practical considerations, such as energy consumption and environmental impact, also warrant attention. Simulating large photonic systems can be resource-intensive, requiring high-performance computing infrastructure that consumes significant energy. Research into energy-efficient algorithms and hardware solutions is an emerging area that aligns with global sustainability goals. Quantum photonic simulation faces a diverse array of challenges, ranging from scalability and noise modeling to user accessibility and ethical considerations. Addressing these challenges will require a combination of technological advancements, interdisciplinary collaboration, and community-driven initiatives. By overcoming these hurdles, the quantum photonics field can unlock new possibilities for research, education, and innovation.

\section{Future Directions}

\subsection{Bridging Simulation and Experimentation}

One of the most critical future directions for quantum photonic simulators is bridging the gap between theoretical simulations and experimental implementations. Current simulators often operate in isolation from the complexities of physical hardware, resulting in discrepancies between simulated and real-world behavior. Future tools must prioritize seamless integration with experimental platforms, enabling researchers to transition smoothly from design to implementation.

Developing simulation frameworks that incorporate real-world imperfections, such as fabrication errors, environmental noise, and thermal fluctuations, is a vital step in this direction. These frameworks should provide interfaces that connect circuit-level designs with device-level models, ensuring that simulation results align closely with experimental realities. Additionally, advancements in photonic chip characterization techniques will play a pivotal role in validating and refining these simulations.

The integration of quantum photonic simulators with hardware control systems also holds significant promise. By connecting simulation tools directly to photonic devices, researchers can perform real-time calibration, adaptive optimization, and feedback-based experiments. These advancements will enable more efficient use of experimental resources and accelerate the iterative process of quantum system development.

\subsection{Scalable Simulation Techniques}

The scalability of quantum photonic simulations remains a persistent challenge, particularly for systems with high levels of entanglement or non-Gaussianity. Future research must focus on developing novel algorithms and representations that enable scalable simulations without sacrificing accuracy.

Hybrid approaches that combine tensor networks, variational methods, and hardware acceleration offer significant potential. Tensor networks can efficiently represent weakly entangled states, while variational methods provide an approximate solution for complex systems. Integrating these techniques with GPU-accelerated solvers or quantum-inspired hardware will further expand the scalability of simulators.

Another promising avenue is the development of adaptive algorithms that dynamically allocate computational resources based on the complexity of the system being simulated. These algorithms can adjust the precision of state representations, the number of sampled photons, or the resolution of phase-space grids, optimizing performance for a given task. Incorporating machine learning into these adaptive algorithms can enhance their efficiency, making scalable simulations more accessible to researchers.

\subsection{Advancing Non-Gaussian Modeling}

Non-Gaussian states and operations are central to universal quantum computation, yet their accurate simulation remains a formidable challenge. Future simulators must prioritize the development of techniques that handle non-Gaussianity more effectively, balancing computational cost with precision.

One area of focus is improving Fock state representations for systems with high photon counts. Current methods scale exponentially with the number of photons, making them impractical for large systems. Sparse matrix techniques, compressed sensing, and other resource-efficient methods could help reduce the memory and computational requirements of these representations \cite{scott2023algorithms}.

Another critical need is the development of better noise models for non-Gaussian systems. Photon loss, dark counts, and detector inefficiencies all disproportionately affect non-Gaussian operations, yet existing models often simplify or overlook these effects. Future simulators should incorporate detailed, experimentally validated noise models that account for the unique challenges of non-Gaussian quantum photonics.

\subsection{Hybrid Quantum-Classical Ecosystems}

Hybrid quantum-classical workflows represent one of the most exciting directions in quantum photonics. These workflows combine the computational power of classical systems with the unique capabilities of quantum photonic circuits, enabling applications such as optimization, machine learning, and quantum chemistry. Future simulators must enhance their support for these workflows, providing robust frameworks for integration and dynamic interaction.

One priority is developing seamless interfaces between quantum simulators and classical programming environments. Tools like Xanadu’s Strawberry Fields have pioneered this approach, integrating with TensorFlow and PyTorch to enable hybrid workflows. Future developments could include support for additional machine learning frameworks, as well as real-time interaction with classical optimization algorithms.

Another focus is expanding the range of hybrid applications supported by photonic simulators. For instance, simulators could incorporate quantum annealing or adiabatic computing models \cite{mc2024towards}, providing a broader toolkit for researchers exploring hybrid quantum-classical solutions. These advancements will make photonic simulators more versatile and relevant to a wide range of fields.

\subsection{Enhanced Visualization and User Interfaces}

Visualization and user interfaces are critical for making quantum photonic simulators accessible to a broader audience. Future simulators must prioritize the development of advanced visualization tools that enable users to intuitively understand complex quantum phenomena.

Real-time visualization of photon dynamics, quantum state evolution, and circuit configurations will be essential. 3D visualizations of waveguides, beam splitters, and photon trajectories can provide researchers with a clearer understanding of circuit behavior, while interactive tools for exploring parameter spaces can accelerate the design process. Visualization frameworks should be modular, allowing users to customize the level of detail and complexity displayed based on their specific needs.

The development of user-friendly interfaces, particularly for educational and outreach purposes, is another critical area. Leveraging modern UI frameworks like SwiftUI \cite{fadda2024ios}, future simulators can offer drag-and-drop circuit design, real-time feedback, and integrated tutorials. These features will lower the entry barrier for new users and foster a broader understanding of quantum photonics.

\subsection{Open-Source Collaboration and Standardization}

Open-source initiatives have been instrumental in advancing quantum simulation tools, but there is still room for growth in collaborative development and standardization. Future efforts should focus on creating unified platforms that bring together the strengths of existing tools while addressing their limitations.

Standardizing data formats, circuit representations, and simulation outputs is a critical step in this process. These standards will enable seamless interoperability between simulators and other software ecosystems, such as machine learning frameworks or physical design tools. Collaborative platforms that allow researchers to contribute plugins, modules, or benchmarks will also accelerate innovation and ensure that simulators evolve in response to the needs of the community.

Open-source simulators must also prioritize accessibility and inclusivity. Providing comprehensive documentation, tutorials, and community support will make these tools more approachable to a diverse range of users, from seasoned researchers to students and educators. By fostering a collaborative and inclusive environment, open-source initiatives can drive the widespread adoption of quantum photonic simulators.

\subsection{Integration with Emerging Technologies}

The field of quantum photonics is evolving rapidly, and simulators must keep pace with emerging technologies to remain relevant. For instance, the integration of quantum photonic simulators with photonic quantum processors, quantum networking platforms, and advanced hardware accelerators is an area of growing interest.

Simulators could also incorporate new paradigms such as topological photonics, which leverages topologically protected states for robust quantum operations. Supporting these technologies will require the development of specialized models and algorithms, as well as close collaboration with experimental researchers.

Another emerging area is the use of quantum-inspired hardware, such as optical neural networks or photonic accelerators, to enhance the performance of the simulators themselves. These technologies have the potential to reduce the computational bottlenecks of classical hardware, enabling more efficient simulations of large photonic systems.

\subsection{Ethical and Practical Considerations}

As quantum photonic simulators become more powerful, ethical and practical considerations will play an increasingly important role in their development. Ensuring that these tools are used responsibly, particularly in applications like cryptography or secure communications, will be a salient challenge. Simulators must include safeguards to prevent misuse and provide transparency about their capabilities and limitations. Practical considerations, such as energy efficiency and environmental impact, are also critical. Simulating large photonic systems requires significant computational resources, which can have a substantial carbon footprint. Research into energy-efficient algorithms and sustainable computing practices will be essential for mitigating these impacts and aligning with global sustainability goals.

\subsection{Democratizing Access to Quantum Photonics}

Finally, the future of quantum photonic simulators lies in their accessibility. By making these tools widely available and easy to use, the community can unlock new opportunities for innovation, education, and collaboration. This democratization will require concerted efforts in interface design, open-source development, and outreach initiatives, ensuring that quantum photonic technologies benefit a broad and diverse audience.

\section{Conclusion}

Quantum photonic simulation has rapidly evolved as a critical component of modern quantum computing, enabling the study and design of photonic circuits essential for quantum information processing, communication, and sensing. This review has examined the transition from Gaussian to non-Gaussian quantum photonic models, the computational challenges associated with simulating large-scale photonic systems, and the state-of-the-art techniques used to address these challenges. Gaussian states provide an efficient computational framework due to their representation in covariance matrices and phase-space formalism, making them highly tractable. However, non-Gaussian states, which are fundamental for universal quantum computation and error correction, introduce significant computational complexity, necessitating advanced numerical techniques such as tensor networks, GPU acceleration, and hybrid quantum-classical algorithms.

A comparative analysis of various quantum photonic simulation platforms, including Strawberry Fields, QuTiP, Piquasso, and Perceval, reveals the diversity of available approaches tailored to different computational needs. These simulators differ in their treatment of continuous-variable (CV) and discrete-variable (DV) quantum systems, the efficiency of their numerical solvers, and their ability to integrate with modern hardware accelerators. Special emphasis was placed on the role of high-performance computing (HPC) architectures, including GPUs, TPUs, and system-on-a-chip (SoC) solutions, in accelerating quantum photonic simulations. The integration of machine learning techniques further enhances the scalability of simulations, allowing for efficient quantum circuit optimization and noise mitigation.

Despite the impressive progress in photonic quantum simulation, several challenges remain. Noise modeling and error mitigation techniques must be further refined to accurately replicate real-world quantum photonic hardware, accounting for effects such as photon loss, dark counts, and mode mismatch. Additionally, as quantum photonic circuits scale, tensor network methods must be optimized to manage computational overhead, and hardware acceleration strategies should be tailored to photonic-specific problems. Future research should focus on improving hybrid quantum-classical approaches, where classical pre-processing and post-processing optimize quantum resource utilization.

The continued advancement of quantum photonic simulation software and hardware co-design will be instrumental in the development of large-scale, fault-tolerant photonic quantum computers. By integrating AI-driven optimization, efficient numerical solvers, and high-performance architectures, quantum photonic simulators will become indispensable tools in pushing the boundaries of quantum information science. The next generation of simulators will need to provide seamless interoperability with experimental platforms, fostering collaboration between theoretical and experimental quantum photonics communities. Addressing these challenges will ensure that photonic quantum computing continues its trajectory toward practical and scalable implementations, paving the way for groundbreaking applications in quantum computing, cryptography, and fundamental physics research.

\section*{Acknowledgment}
This research was not funded. Any opinions, findings, conclusions, or recommendations expressed in this review are those of the author(s) and do not necessarily reflect the views of their respective affiliations.

\section*{Conflicts of interest}
The author(s) declare no competing interests.

\nomenclature{DV}{Discrete-Variable}
\nomenclature{CV}{Continuous-Variable}
\nomenclature{VQAs}{Variational Quantum Algorithms}
\nomenclature{QuTiP}{Quantum Toolbox in Python}
\nomenclature{MPS}{Matrix Product State}
\nomenclature{HPC}{High-Performance Computing}
\nomenclature{TPUs}{Tensor Processing Units}
\nomenclature{ML}{Machine learning}
\nomenclature{2D}{Two-dimensional}
\nomenclature{3D}{Three-dimensional}
\nomenclature{SoC}{System on a Chip}
\nomenclature{AI}{Artificial Intelligence}
\nomenclature{ML}{Machine Learning}
\nomenclature{CUDA}{Compute Unified Device Architecture}
\nomenclature{FDTD}{Finite-Difference Time-Domain}
\nomenclature{UI}{User Interface}
\nomenclature{FPGA}{Field Programmable Gate Arrays}
\nomenclature{API}{ Application Programming Interface}
\nomenclature{IEEE}{Institute of Electrical and Electronics Engineers}
\nomenclature{SPDC}{Spontaneous Parametric Down-Conversion}
\nomenclature{IPEA}{International Preliminary Examination Authority}
\nomenclature{SPOPO}{Synchronously Pumped Optical Parametric Oscillator}
\nomenclature{NC}{Nonlinear Crystal}
\nomenclature{BS}{Beam Splitter}
\nomenclature{SPD}{Single-photon Detector}
\nomenclature{MHz}{Megahertz}
\nomenclature{PBS}{Polarizing Beam Splitter}
\nomenclature{BBO}{Barium borate}
\nomenclature{HWP}{Half-wave Plate}
\nomenclature{HWHM}{half-width at half-maximum }
\nomenclature{SiPM}{Silicon Photomultiplier}
\nomenclature{DCR}{Dark Counts Rate}
\nomenclature{TFLOPS}{Tera Floating-point Operations per Second}
\nomenclature{PQC}{Photonic Quantum Computing}
\printnomenclature

\bibstyle{sn-mathphys-num}

\end{document}